\documentclass[osajnl,twocolumn,showpacs,superscriptaddress,10pt]{revtex4-1} 
\usepackage{color}
\usepackage{xspace}
\newcommand{\genunit}[2]{\ensuremath{#1~\text{#2}}\xspace}
\newcommand{\cm}[1]     {\genunit{#1}{cm}}
\newcommand{\kev}[1]    {\genunit{#1}{keV}}
\usepackage{amsmath,amssymb,graphicx}
\usepackage{epstopdf}
\usepackage{float}
\begin{document}

\title{Development of an Ideal Observer that Incorporates Nuisance Parameters and Processes List-Mode Data }

\author{Christopher J. MacGahan}\email{Corresponding author:cmacgahan@optics.arizona.edu}
\affiliation{College of Optical Sciences, The University of Arizona, 1630 E. University Blvd, Tucson, AZ, 85721, USA}
\affiliation{Sandia National Labs, Livermore, CA 94551}
\thanks{\copyright 2016. Optical Society of America. One print or electronic copy may be made for personal use only. Systematic reproduction and distribution, duplication of any material in this paper for a fee or for commercial purposes, or modifications of the content of this paper are prohibited.}

\author{Matthew A. Kupinski}
\affiliation{College of Optical Sciences, The University of Arizona, 1630 E. University Blvd, Tucson, AZ, 85721, USA}

\author{Nathan R. Hilton}
\author{Erik M. Brubaker}
\author{William C. Johnson}
\affiliation{Sandia National Labs, Livermore, CA 94551}

\begin{abstract}
Observer models were developed to process data in list-mode format in order to perform binary discrimination tasks for use in an arms-control-treaty context. Data used in this study was generated using GEANT4 Monte Carlo simulations for photons using custom models of plutonium inspection objects and a radiation imaging system.  Observer model performance was evaluated and presented using the area under the receiver operating characteristic curve. The ideal observer was studied under both signal-known-exactly conditions and in the presence of unknowns such as object orientation and absolute count-rate variability; when these additional sources of randomness were present, their incorporation into the observer yielded superior performance. 
\end{abstract}

\maketitle

\section{Introduction}
Nuclear imaging systems used to reconstruct images of treaty-accountable items (TAI) have been proposed as a component of treaty verification between countries. A disadvantage to the imaging techniques is that the monitor can use knowledge of the detector data and imaging system to reconstruct sensitive geometrical information on the TAI. A physical or software information barrier (IB) must then be used to prevent the disclosure of sensitive information to unauthorized individuals or governments. An example of an IB is the CIVET system developed by Brookhaven National Laboratory \cite{CIVET}, which puts a high-resolution gamma spectrometer behind an IB. Both the host and monitor can authenticate the device, but the monitor cannot access the sensitive gamma spectra; it only sees a final decision. This makes the device expensive to develop and authenticate, for reasons noted in the above cited paper, and ultimately reduces confidence in the verification results. For additional examples of information barriers, we point the reader to Sandia National Laboratories' TRIS and TRAD systems \cite{TRIS, TRAD, TRISTRADREVIEW}.

To overcome the need for an information barrier, we have developed mathematical models (called observer models in this work) to classify unverified test objects. The observer models are built on acquired calibration data from a trusted TAI and identify tested sources using raw projection data. We focused on the development of models that process list-mode (LM) events---which we define as the interaction of a particle in the detector. The use, and subsequent disposal, of LM events as they are acquired by the system means that an ``image'' is never actually formed by the imaging system either in terms of a projection image or a reconstruction of the object. LM processing is essential to any treaty-verification system that does not need an IB. While these models have been developed to perform the unique task of treaty verification, LM observers could also be extended for use in security scanning---offering the possibility of identifying threats without revealing the intimate details of the objects themselves.

In recent years, a multitude of new classification methods have been developed, such as tree classifiers \cite{TreeSurvey}, support vector machines (SVM)\cite{SVM}, and artificial neural networks \cite{NNMethodology}. While the authors have not done a thorough analysis of these models, it appears difficult to adapt some of these routines to process LM data, which is a very constricting condition. Random forest classifiers \cite{RFBreiman,RFLiaw} operate by constructing many random decision trees (where decisions could consist of a comparison of one of the data variables with a number) from bootstrapped samples of calibration data and then aggregating the tree results to make a decision. This classifier requires the aggregation of testing data to make decisions and such information would necessitate an IB. Likewise, a non-linear SVMs would require knowledge of the complete data set. Only a linear SVM would satisfy the LM requirement. Similarly, artificial neural networks cannot process LM data because they use a nonlinear sigmoidal function after each node.  The best any of these methods could do is to approximate the Bayesian ideal observer \cite{Foundations}, which provides an upper bound on performance. 
 
In this paper we choose to directly calcuate the Bayesian ideal observer, which has complete probabilistic knowledge of the image data and can be adjusted to process LM data. By its nature, it offers the best possible performance for a given task. This comes at the cost of storing a significant amount of sensitive calibration data. We do not view this model as a practical solution to the IB dilemma (in fact the requirement of complete probabilistic knowledge makes the ideal observer difficult to create in practice), but as a case study of the optimally performing LM observer under ideal conditions. This is the first time the ideal observer has been used with LM data to perform binary-discrimination tasks for the purpose of hiding sensitive information. This observer will provide an upper bound to gauge the performance of future LM observer models that store less information. In addition, we present results of marginalization over nuisance parameters such as unknown orientation and source activity and show that inclusion of these nuisance parameters in our model leads to improved performance and is often necessary. 
 
We previously developed a signal-known-exactly (SKE) ideal observer that processes LM data~\cite{2014NSSCP}.  The SKE ideal observer assumes precise knowledge of the objects being imaged (the signals) such as their orientation, position, count rate, etc.  When these parameters are not known exactly, we refer to them as nuisance parameters, as they prevent correct discrimination of objects but are not of concern to the task itself. This paper generalizes the ideal-observer theory to include the impact of these nuisance parameters.

\section{Theory}
Before beginning discussion of the observers, it is useful to formalize the theory for LM data. Much of this notation is taken from work developed by Barrett, Parra, and Caucci~\cite{LMParra, LMCaucciOA}. Additional discussion on list-mode theory and its applications can be found in work by Clarkson \cite{clarkson2012asymptotic} and Jha \cite{jha2013analytic}. It is worth mentioning that there is a fundamental difference in the motivations for their work and this work. While they utilize LM data to prevent the loss of information that comes with binning data, our desire is to discriminate sources with LM data to prevent the aggregation of information that would necessitate an IB. 

There are two components to the image data collected: the number $N$ of photons and neutrons that interact with the detector, which is Poisson distributed, and the LM data attributes $A_n$ estimated for each detected event.  
Each $A_n$ contains all of the detectable information for the $n^{th}$ observed particle. For a coded-aperture imager (to be discussed later) this data can be defined as
\begin{equation}\label{eqn:LMdef}
	A_n=\{\text{particle type, pixel number, energy deposited}\},
\end{equation}
where $n$ goes from $1$ to $N$. Though we use a specific imager in our experiments, the methods developed here should apply regardless of the chosen detection system as long as the system can generate LM data. 

Nuisance parameters are potential sources of variability that affect the data acquired, but are not of interest for performing the task. Incorporation of nuisance parameters into our observer model helps compensate for the performance losses these unknowns introduce. Some nuisance parameters, such as object orientation or location, apply to all objects being imaged. Other nuisance parameters, such as variations in gamma or neutron intensities or energy distributions, which may be caused by unknown material compositions or ages, could be object-dependent. We define $\gamma_j$ as the set of of nuisance parameters, e.g., 
\begin{equation}\label{eqnNuisanceParamdef}
	\gamma_j=\{\text{object orientation, object location, source age}\},
\end{equation}
where the index $j$ is used to denote the object type. In this work $j$ can be either $1$ or $2$ since we are considering discrimination tasks between two object classes.

The ideal observer~\cite{Foundations} is defined as
\begin{equation}\label{eqn:IdOb}
\Lambda(\{A_n\},N)=\frac{pr(\{A_n\},N|H_2)}{pr(\{A_n\},N|H_1)},
\end{equation}
where $pr(\cdot)$ denotes a probability density function (pdf), $H_2$ denotes the hypothesis that object 2 is being imaged, and $H_1$ the hypothesis that object 1 is being imaged. We distinguish between $pr(\cdot)$ and a discrete probability function $Pr(\cdot)$ with capitalization, though we recognize that in \eqref{eqn:IdOb}, the arguments are a mixture of discrete and continuous random variables and we chose to use the $pr$ notation in these cases. The ideal observer thresholds the likelihood ratio to make decisions and declare the data from class 1 or class 2.  Note that the likelihood includes the LM data as well as the number of detected events $N$, which is not LM data, as it requires accumulating information (the event count) over many events.  Though not explicitly stated in the above equation, the likelihoods and ideal observer developed in this work are dependent on acquisition time.

\subsection{Signal-Known-Exactly Ideal Observer}

The SKE ideal observer assumes that the nuisance parameters are known and thus the likelihood ratio is given by
\begin{equation}\label{eqn:IdObSKE}
  \Lambda_{SKE}(\{A_n\},N|\gamma_1,\gamma_2)=\frac{pr(\{A_n\},{N}|\gamma_2,H_2) }{ pr(\{A_n\},{N}|\gamma_1,H_1) }.
\end{equation}
We start by developing a LM form for the SKE likelihoods used in \eqref{eqn:IdObSKE}. This derivation will closely follow a prior discussion of the subject~\cite{2014NSSCP}.  Under a known set of nuisance parameters the likelihood that the data $(\{A_n\},{N})$ is due to object $j$ is
\begin{equation}\label{eqn:SKELikelihoodNuisance1}
pr(\{A_n\},{N}|\gamma_j,H_j)=pr(\{A_n\}|{N}, \gamma_{j},H_j) Pr({N}|\gamma_{j},H_j).
\end{equation}
The first term is a continuous probability density on observing some set of LM data given full knowledge of the source nuisance parameters. The second term is a Poisson probability on the number of counts observed. As each event is independent, \eqref{eqn:SKELikelihoodNuisance1} can be written as
\begin{equation}\label{eqn:SKELikelihoodNuisance2}
pr(\{A_n\},{N}|\gamma_j,H_j)= Pr(N|\gamma_{j},H_j) \prod_{n=1}^{N} pr(A_{n}|\gamma_{j},H_j), 
\end{equation}
where the last term $pr(A_{n}|\gamma_{j},H_j)$ is the probability of observing the LM event data $A_n$  given that object $j$ is being imaged and with  known nuisance parameters $\gamma_j$.  

The Poisson probabilities depend on the mean count rate for events originating from the object being imaged (the source) $\overline{N}^{(s)}_j$ and the mean count rate for events originating outside the object (called background events) $\overline{N}^{(b)}$, both of which depend on the set of nuisance parameters $\gamma_j$.   We define the overall mean count rate for hypothesis $H_j$ as $\overline{N}_j=\overline{N}^{(s)}_j+\overline{N}^{(b)}$.  In \eqref{eqn:SKELikelihoodNuisance1} we can replace $Pr(N|\gamma_j,H_j)$ with a Poisson probability with mean $\overline{N}_j$, i.e., $Pr(N|\overline{N}_j)$.

To make sense of the second term, $pr(A_{n}|\gamma_{j},H_j)$, we next define a variable $h_n$ that describes the origin of the $n^{th}$ detected particle. To do so, we include the probability that an event is a background event ($h_n=h^{(b)}$) or a source event ($h_n=h^{(s)}$). Including these conditional probabilities in our LM term,
\begin{equation}
\label{eqn:SignalBackgroundsplit}
\begin{split}
&pr(A_n|\gamma_j,H_j) = \\
&pr(A_n|\gamma_j,h_n=h^{(b)})Pr(h_n=h^{(b)}|\gamma_j,H_j) + \\
&pr(A_n|\gamma_j,H_j,h_n=h^{(s)})Pr(h_n=h^{(s)}|\gamma_j,H_j),
\end{split}
\end{equation}
where $Pr(h_n=h^{(b)}|\gamma_j,H_j)$ is the probability that the detected event came from the background, and $Pr(h_n=h^{(s)}|\gamma_j,H_j)=1-Pr(h_n=h^{(b)}|\gamma_j,H_j)$ the probability that the detected event originated from the source. These probabilities are equal to the ratio between the mean number of background or signal counts and the total mean number of counts. The dependence of the LM data for a background event on $H_j$ was dropped because the background distribution will be the same for either object being imaged.

We have thus far expanded the likelihoods found in the numerator and denominator of \eqref{eqn:IdObSKE} to include a Poisson term on the total detected counts, terms which account for the probability of a background or a source event (another nuisance parameter related to the count rate), and the probability density of observing LM data for a source event and a background event.  These last two probabilities include the distribution of where an event will occur in the detector (i.e., the imaging aspect of the system) as well as the distribution associated with the energy of the event (i.e., the spectral aspect of the system).  These distributions must be determined either through calibration or through Monte Carlo simulations; we discuss our method for calculating these distributions later in this paper.

Replacing the likelihoods in \eqref{eqn:IdObSKE} with the discussed expressions reveals that
\begin{equation}\label{eqn:IdObNuisanceSKE1}
\begin{split}
\Lambda_{SKE}(\{A_n,\},N&|\gamma_{1}, \gamma_{2})=\\
&\frac{Pr(N|\overline{N}_2)}{Pr(N|\overline{N}_1)}\prod_{n=1}^N  \frac{pr(A_n|\gamma_2,H_2)}
{pr(A_n|\gamma_1,H_1)}.
\end{split}
\end{equation}
Both the numerator and denominator inside the product utilize the background and source decomposition shown in \eqref{eqn:SignalBackgroundsplit}.  

Current implementation of this SKE ideal observer occurs in two stages. First, in the calibration stage, a pair of high-statistics simulated LM detector data sets are generated for two different sources. These data sets are used to find $\overline{N_j}$ and are binned by energy and pixel number to find a probability density on observing the LM data $pr(A_n|\gamma_j,H_j)$. In practice, the observer model would then be evaluated on different training data sets, yielding probability distributions on the test statistic for each of the two objects. A threshold would then be chosen for optimal decision making. 

Second, in the testing stage, the test statistic $\Lambda$ is initialized to one. For each detected event, the test statistic is multiplied by the ratio of observing that event's data $A_n$ given the two hypotheses and the known nuisance parameters. That data $A_n$ is then forgotten. At the end of the acquisition time, $\Lambda$ is multiplied by the ratio of the probabilities for observing $N$ counts under the two hypothesis. Note that the recording of $N$ particles is not strictly LM data. Finally, $\Lambda$ is thresholded to make a decision. 

The choice to compute the test statistic rather than the individual likelihoods was made for computational ease. The LM probabilities used in this model are often very small (the smaller the bin size, the lower the probability values will be). Calculation of an individual likelihood expression such as $pr({A_n},N|\gamma_1,H_1)$ is difficult as it is the product of small numbers and goes to zero computationally after a handful of list-mode events. This can be overcome through the use of log likelihoods, or by tracking only the test statistic. We chose to just track the test statistic, multiplying it by the probability ratio of observing a given detected particle for each of the two hypothesis. This ratio is generally close to one, allowing a large number of events can be processed.

Finally, we note that the SKE ideal observer requires storage of both spatial and spectral information. In our treaty-verification application, this information could be used to determine sensitive isotopic and spatial information about the object. Therefore, storage of this calibration data would need to be behind an IB. 

\subsection{Ideal Observer Incorporating Nuisance Parameters}
When the nuisance parameters $\gamma_1$ and $\gamma_2$ are not known exactly, the likelihood expressions in the ideal observer must be integrated over the probability densities of those nuisance parameters. \eqref{eqn:IdOb} becomes   
\begin{equation}\label{eqn:IdObNuisance}
  \Lambda(\{A_n\},N)=\frac{\int\,pr(\{A_n\},{N}|\gamma_2,H_2) pr(\gamma_2) \mathrm{d}\gamma_2}{ \int\,pr(\{A_n\},{N}|\gamma_1,H_1) pr(\gamma_1) \mathrm{d}\gamma_1}.
\end{equation}
The probability density on the LM data $\{A_n\}$ is conditioned on knowledge of the nuisance parameters and then averaged over the distribution of the nuisance parameters. This expression can be evaluated in different ways depending on the form the likelihood takes and the number of nuisance parameters integrated over.  

Monte Carlo sampling of the integrals in the numerator and denominator offers the most straightforward solution, but proves difficult in instances where the numerator and denominator are both very close to zero computationally. This can be circumvented by using the log likelihood. For a sampled $\gamma_{j,n}$ (where n goes from 1 to the number of Monte Carlo samples) from a distribution $pr(\gamma_j)$, the value $log(pr(\{A_n\},{N}|\gamma_{j,n},H_j))$ can be calculated. To proceed without the storage of LM data, all samples of $\gamma_j$ must be taken before the data is processed, and each log likelihood expression updated as the LM data is read in. At the end, a common factor is subtracted from both numerator and denominator of \eqref{eqn:IdObNuisance}, and the terms are re-exponentiated and added. This procedure makes the problem computationally feasible. In testing, the method generally requires storage of the individual likelihood values $pr(\{A_n\},{N}|\gamma_j,H_j)$ for each of the two sources and chosen samples of $\gamma_j$.

The incorporation of nuisance parameters brings a practical concern. The dependence of the true detected spatial and energy distributions on certain nuisance parameters is complex. In order to use the ideal observer in practice, the host country would need to image their TAIs under many different conditions in order to properly train the ideal observer. While the nuisance parameter priors discussed in this section were continuous, in reality they would likely need to be treated as discrete. In addition, the host would need to investigate the objects themselves to properly identify the prior distributions. Any deviation between the prior distributions chosen by the host and the distribution of those nuisance parameters when the monitor is testing the objects will degrade performance. 

Monte Carlo integration proved sufficient to evaluate the likelihoods for the experiments we chose to perform in this paper as only one nuisance parameter was treated at a time. As the dimensionality of the nuisance parameters $\gamma_1$ and $\gamma_2$ increases, evaluation of this integral through standard Monte Carlo methods becomes difficult due to slow convergence \cite{MCSims}. This can be improved through Quasi-Monte Carlo methods, but a faster technique to calculate this integral is Markov-Chain Monte Carlo~\cite{MCMC}. Markhov-Chain Monte Carlo integration continuously resamples the nuisance parameters based on a proposal density $pr(\gamma_{j,new}|\gamma_{j,old})$, accepting the new evaluation when $pr({A_n},N|\gamma_j,H_j)$ increases and rejecting the step with a certain probability when the likelihood decreases. To use this method while keeping the LM requirement, the probability of observing our data would need to be calculated over a predefined grid on the nuisance parameter values. The proposal density on the sampled nuisance parameter values would also need to be discrete due to the argument made in the prior paragraph.

In the following section, we derive another method for evaluating \eqref{eqn:IdObNuisance} using posterior probabilities on the nuisance parameters, with its own advantages and disadvantages. 

\subsection{Ideal Observer Using Posterior Probability Density}\label{sec:postpdf}
The following is similar to a derivation by Kupinski in a previous paper~\cite{MCMCKupinski}. Before beginning, we make a slight notational change for convenience. The symbol $\gamma_0$ is now the set of nuisance parameters shared by source 1 and 2, such as variability in background activity, orientation, and location. The symbols $\gamma_1$ and $\gamma_2$ will now be used to describe the nuisance parameters unique to sources 1 and 2, such as the material composition of each source if not exactly known. \eqref{eqn:IdObNuisance} now becomes
\begin{equation}\label{eqn:IdObNuisancePost}
\begin{split}
 &\Lambda(\{A_n\},N)=\\
&\frac{\int\int\,pr(\{A_n\},{N}|\gamma_0,\gamma_2,H_2) pr(\gamma_0) pr(\gamma_2) \mathrm{d}\gamma_0\mathrm{d}\gamma_2}{\int\int\,pr(\{A_n\},{N}|\gamma_0,\gamma_1,H_1) pr(\gamma_0) pr(\gamma_1)\mathrm{d}\gamma_0 \mathrm{d}\gamma_1}.
\end{split}
\end{equation}
Beginning with \eqref{eqn:IdObNuisancePost}, we simplify the denominator back to $pr(\{A_n\},N|H_1)$ and marginalize the numerator over the remaining nuisance parameters $\gamma_1$,
\begin{equation}\label{eqn:IdObNuisancePost1}
\begin{split}
\Lambda(\{A_n\},N)=&\frac{1}{pr(\{A_n\},N|H_1)}\\
&\times \int\int\int pr(\{A_n\},N|\gamma_0,\gamma_2,H_2)... \\
&\;\;\;\;\; pr(\gamma_0) pr(\gamma_1) pr(\gamma_2)  \mathrm{d}\gamma_0 \mathrm{d}\gamma_1 \mathrm{d}\gamma_2.
\end{split}
\end{equation}
Next, multiply both the numerator and the denominator inside the integral by the SKE likelihood for source 1, $pr({A_n},N|\gamma_0,\gamma_1,H_1)$, and acknowledge that $pr({A_n},N|\gamma_0,\gamma_2,H_2)/pr({A_n},N|\gamma_0,\gamma_1,H_1)$ for a specific $\gamma_1$ and $\gamma_2$ is the SKE observer as in \eqref{eqn:IdObNuisanceSKE1},
\begin{equation}\label{eqn:IdObNuisancePost2}
\begin{split}
\Lambda(\{A_n\},N)=&\frac{1}{pr(\{A_n\},N|H_1)}\\
&\times \int\int\int \Lambda_{SKE}(\{A_n\},N|\gamma_0,\gamma_1,\gamma_2)\\
& pr(\{A_n\},N|\gamma_0,\gamma_1,H_1) pr(\gamma_0) pr(\gamma_1) \\
& pr(\gamma_2) \mathrm{d}\gamma_0 \mathrm{d}\gamma_1 \mathrm{d}\gamma_2.
\end{split}
\end{equation}
Next we simplify further using Bayes' rule, creating a posterior probability density,
\begin{equation}\label{eqn:IdObBayes}
\begin{split}	
pr(\gamma_0,\gamma_1|\{A_n\},N,&H_1)=\\
&\frac{pr(\{A_n\},N|\gamma_0,\gamma_1,H_1)pr(\gamma_0)pr(\gamma_1)}{pr(\{A_n\},N|H_1)}.
\end{split}
\end{equation}
Substituting \eqref{eqn:IdObBayes} into \eqref{eqn:IdObNuisancePost2}, we arrive at the final result,
\begin{equation}\label{eqn:IdObNuisancePost3}
\begin{split}
\Lambda(\{A_n\},N)=&\int\,\Lambda_{SKE}(\{A_n\},N|\gamma_0,\gamma_1,\gamma_2)pr(\gamma_2)\\
&pr(\gamma_0,\gamma_1|\{A_n\},N,H_1)\mathrm{d}\gamma_0 \mathrm{d}\gamma_1 \mathrm{d}\gamma_2.
\end{split}
\end{equation}

This integral can be evaluated by sampling the nuisance parameters $\gamma_0$ and $\gamma_1$ from the posterior density $pr(\gamma_0,\gamma_1|\{A_n\},N,H_1)$ and the $\gamma_2$ nuisance parameters from their respective probability densities and performing Monte Carlo integration. This provides an advantage over \eqref{eqn:IdObNuisance} because the SKE ideal observer is generally more computationally feasible than the individual likelihoods and doesn't require the use of log likelihoods. Another advantage to this method is that the posterior pdf provides a distribution on $\gamma_0 , \gamma_1$ values more consistent with the data $(\{A_n\},N)$ than a simple Monte Carlo sample. The integral should therefore converge faster. 

However, the posterior pdf is different for each data set $(\{A_n\},N)$. Computing the posterior pdf in \eqref{eqn:IdObBayes} requires evaluating the likelihood $pr(\{A_n\},N|\gamma_0,\gamma_1,H_1)$ over a large number of points on the $\gamma_1,\gamma_0$ nuisance parameter grid without knowing in advance what $(\{A_n\},N)$ are in order to avoid storage of the LM data. In the end, we are uncertain this method will provide a significant speed increase as a similar number of grid points may be required to evaluate \eqref{eqn:IdObNuisancePost3} as samples needed to effectively evaluate \eqref{eqn:IdObNuisance}.

Regardless of the approach chosen, the ideal observer would need to store many sets of calibration data under different object configurations in order to incorporate nuisance parameters, increasing storage requirements and complexity of operations behind an IB.


\section{Simulation}
\subsection{Objects and Imager}
\begin{figure}
\begin{center}$
\begin{array}{cc}
\includegraphics[width=0.45\columnwidth]{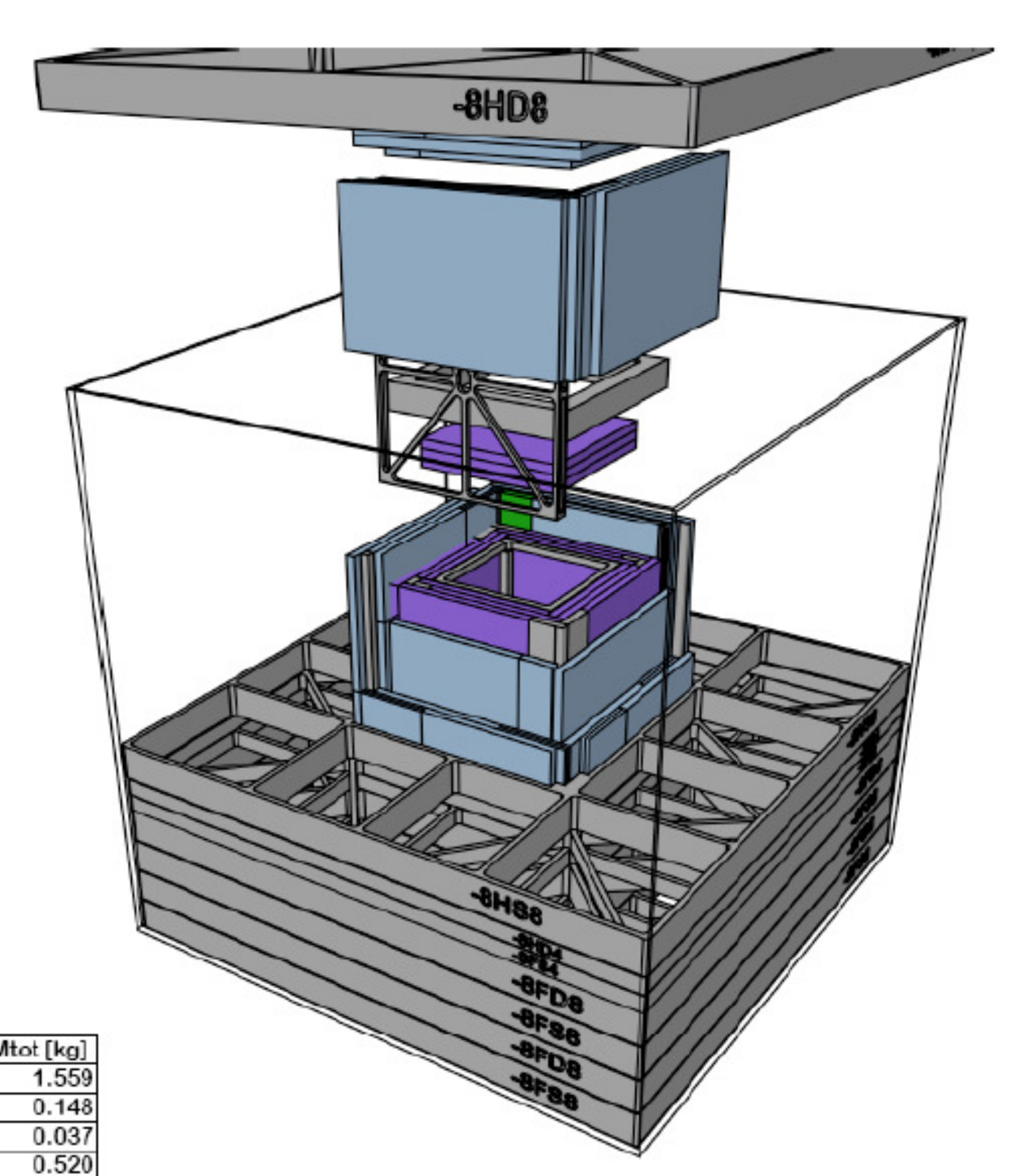} \;\;&\;\;
\includegraphics[width=0.45\columnwidth]{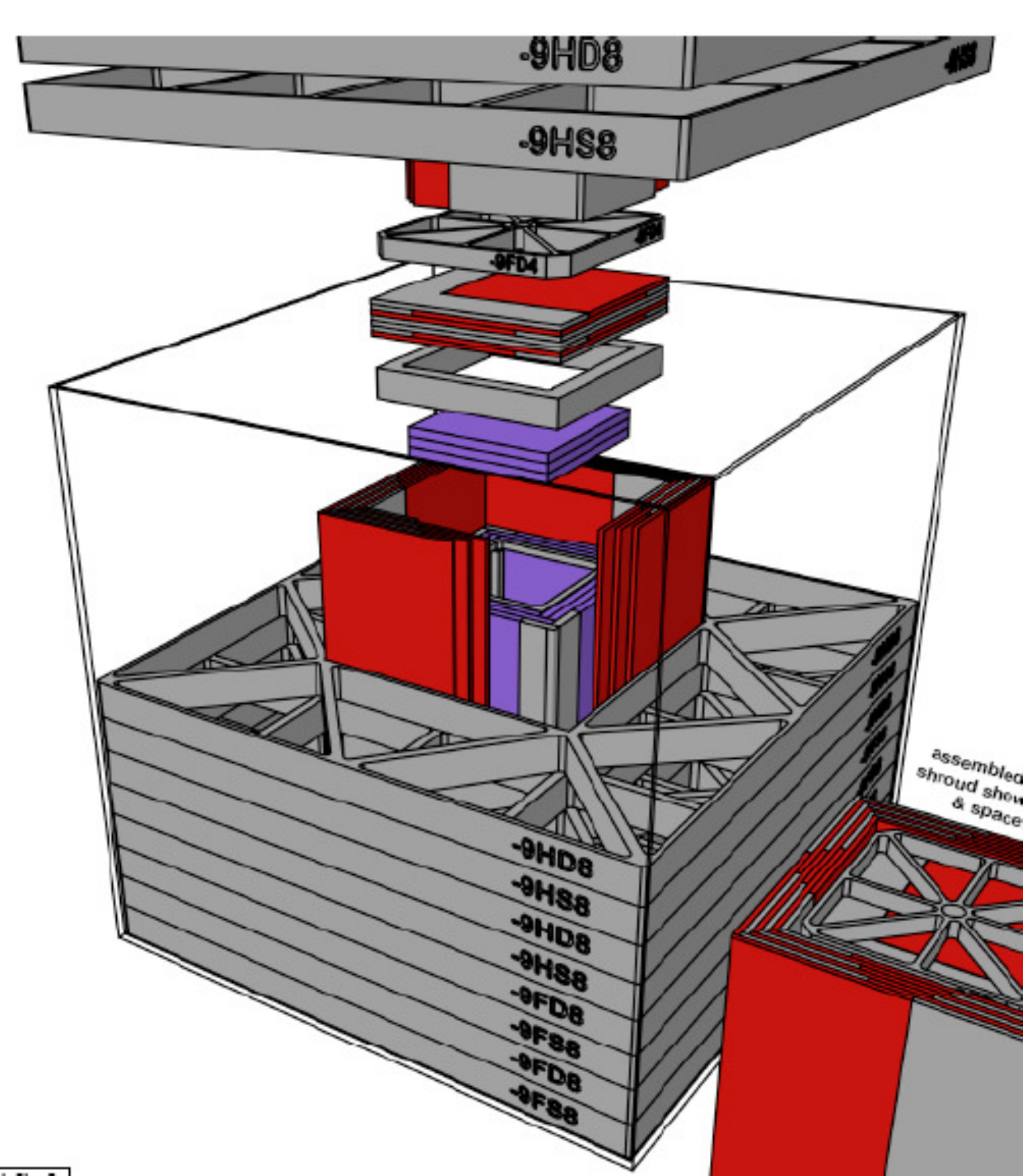}
\end{array}$
\end{center}
\caption{IO8 and IO9 developed by INL~\cite{INL}. IO8 is plutonium shielded by depleted uranium (DU) while IO9 is plutonium shielded by highly-enriched uranium (HEU). Both assemblies are supported by an aluminum framework inside an $8"\times8"\times8"$ aluminum box that is 1" thick.}
\label{fig:IO89specs}
\end{figure}

\begin{figure}
	\centerline{\includegraphics[width=.9\columnwidth]{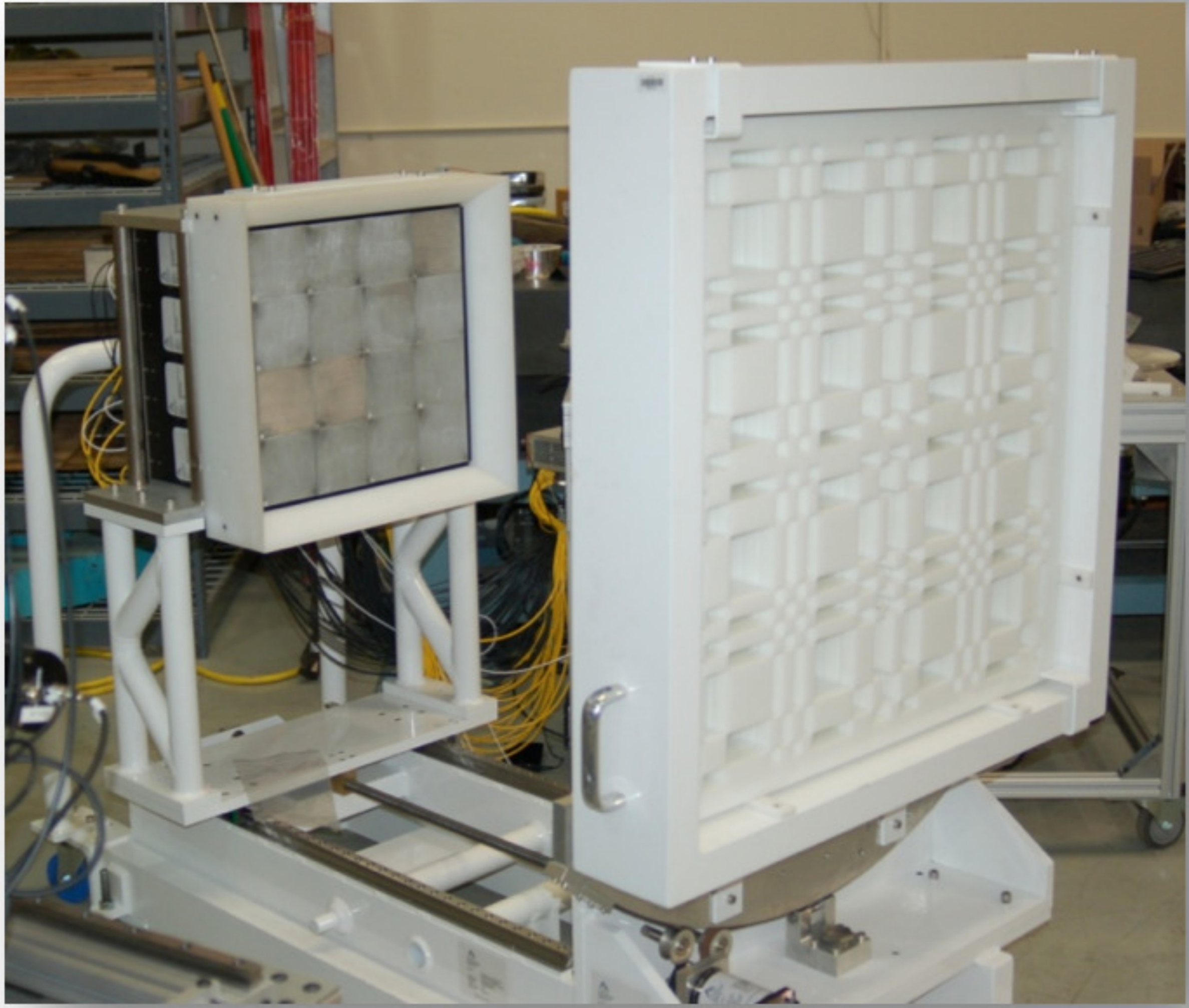}}
	\caption{Fast-neutron coded-aperture imaging system. The imager uses a polyethylene coded aperture and a $4\times4$ array of liquid-scintillator detectors, each consisting of $10\times10$ $(1~\text{cm})^2$ pixels.
A quarter-inch lead plate (not pictured) is positioned in front of the pixelated detectors, blocking low-energy gammas.}
	\label{fig:FNdetector}
\end{figure}

Binary classification tasks were performed using inspection objects developed by Idaho National Laboratory (INL)~\cite{INL}; this paper uses inspection objects labeled 8 and 9 (see Fig. \ref{fig:IO89specs}), which are referred to here as IO8 and IO9. These objects differ in their shielding material, causing a difference in the gamma spectra in the image data. The simulated detector is a fast-neutron coded-aperture imager (see Fig. \ref{fig:FNdetector}), developed by Oak Ridge National Laboratory and Sandia National Laboratories~\cite{FNDetector}.  Though this detector is designed to observe neutrons, it also serves as a low-resolution gamma detector.

\subsection{Forward Model in GEANT4}\label{sec:ForwardModel}
\begin{figure}
	\centerline{\includegraphics[width=.9\columnwidth]{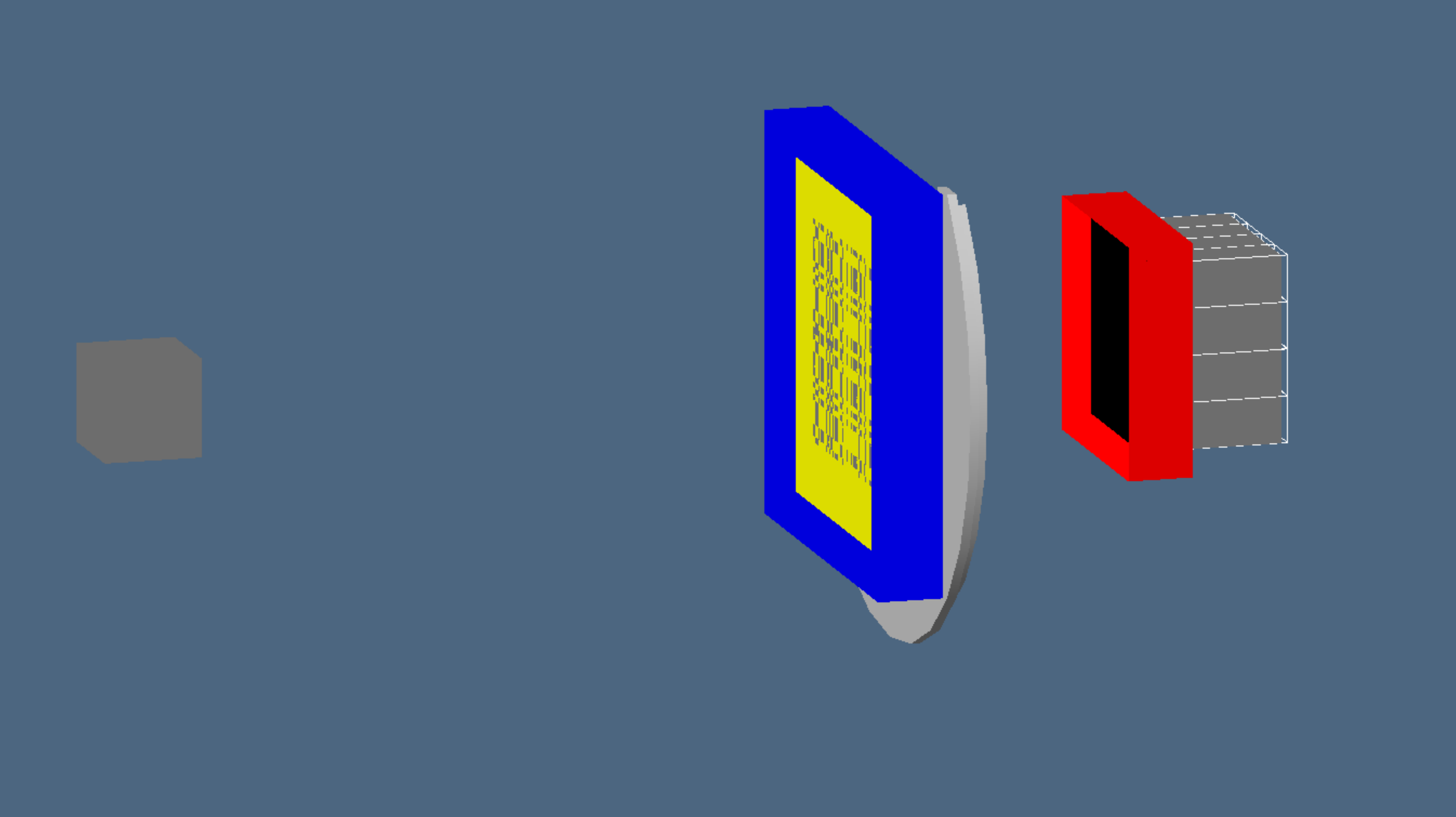}}
	\caption{Geant4 model of system. IO8 is stored inside aluminum cube on the left (grey). The polyetheylene mask is shown in yellow and the grey geometries in the mask are holes. On the right is the detector.}
	\label{fig:G4fullmodel}
\end{figure}

We developed a Monte Carlo transport application that uses the GEANT4 toolkit~\cite{G4Sim,G4Dev} to model gammas emitted by the source objects, as well as the detector response to the emitted radiation. Due to similarities in the geometry between the two objects, neutrons were not incorporated in this study as the spectral and spatial detector data differences between the two studied objects were expected to be minimal. A linear energy bias, as well as a low-energy cutoff of \kev{100} were used to make the simulations computationally feasible. A detector-response code collects the light output, applies an energy smear specific to the detector, and bins it into a mean pixel location; a perfect pulse-shape discrimination between gammas and neutrons was assumed. Visualization of our GEANT4 simulation is shown in Fig. \ref{fig:G4fullmodel}.

To simulate object orientations, we used a method developed by Arvo~\cite{Arvorotation} that uses three random numbers between zero and one to generate random rotations of an object. The object is first rotated a random amount around the $\hat{z}$  axis using the first number; then the $\hat{z}$ axis is rotated to a random location in $\phi, \theta$ space using the last two. To generate stratified samples, we used evenly spaced values of the three random numbers stated in Arvo's work.  Three initial rotations around $\hat{z}$ were chosen. Then $\hat{z}$ was rotated into twenty different points (five in $\phi$, four in $\theta$) on the sphere for sixty total orientations. In the studies in the experiments section, a stated rotation number $x_1x_2x_3$ corresponds to Arvo random numbers $x_1/3$, $x_2/5$, $x_3/4$. As an example, Arvo rotation 111 corresponds to Arvo random numbers $1/3$,$1/5$, and $1/4$.

\begin{figure}[htbp]
	\centerline{\includegraphics[width=.9\columnwidth]{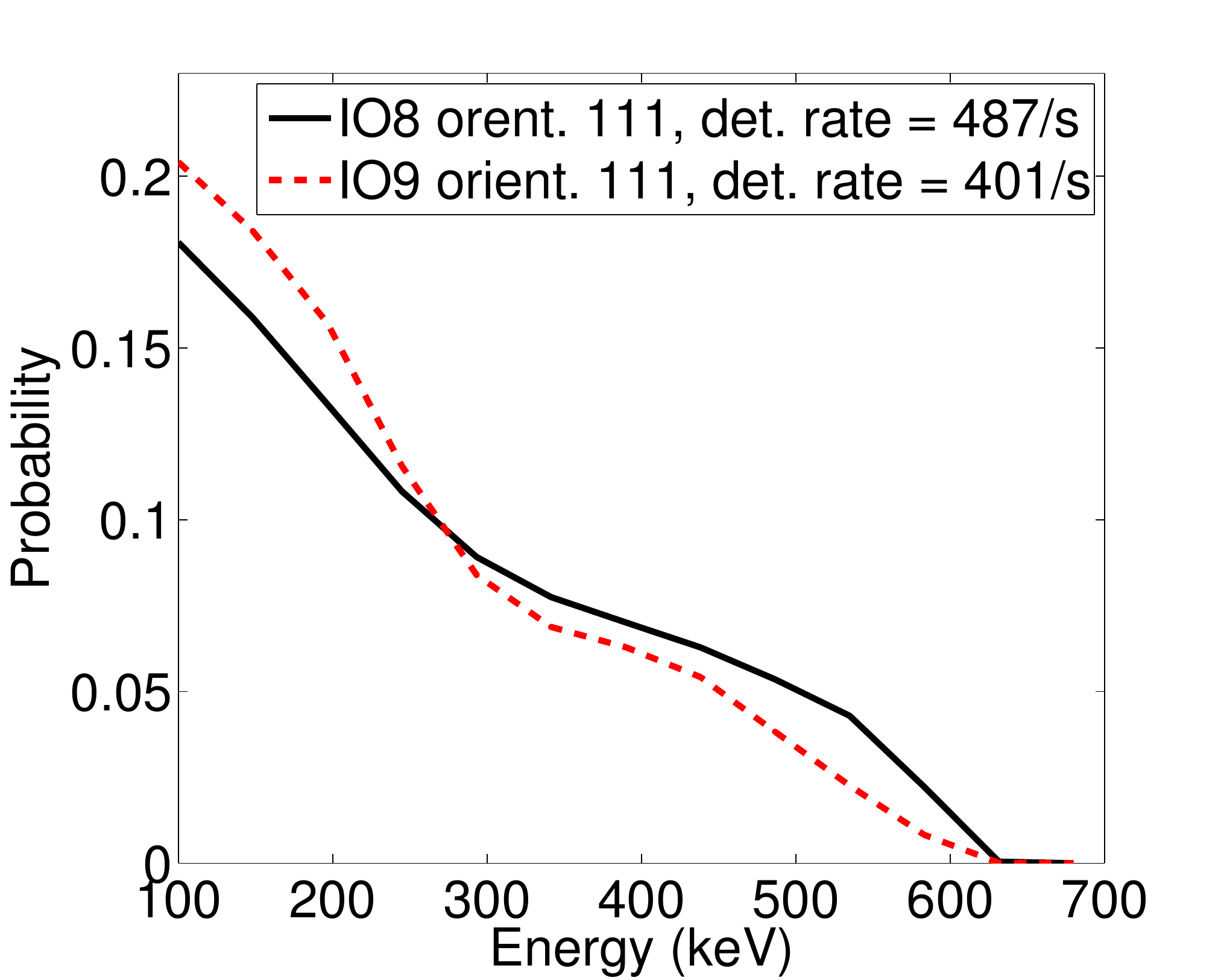}}
	\caption{Comparison of IO8 and IO9 gamma spectra for an off-axis orientation (Arvo numbers 111) that is representative of the majority of orientations of the objects. IO9 energy spectrum is shifted towards lower energies with the more active HEU shielding material.}
	\label{fig:IO8v9GammaSpectra}
\end{figure}

The gamma-ray detection rates and energy spectra present the most significant difference between the two sources because of the shielding components (see Fig. \ref{fig:IO8v9GammaSpectra}) The plotted spectra in this figure are for Arvo numbers 111. The spectral disparity is due to the difference in shielding material, as HEU has a high intensity at \kev{186} and DU a moderate intensity at \kev{1001}. Both the detected spectra and count rate are dependent on the orientation chosen, though the count rate varies more significantly with orientation.

\subsection{Background}
\begin{figure}[htbp]
	\centerline{\includegraphics[width=.9\columnwidth]{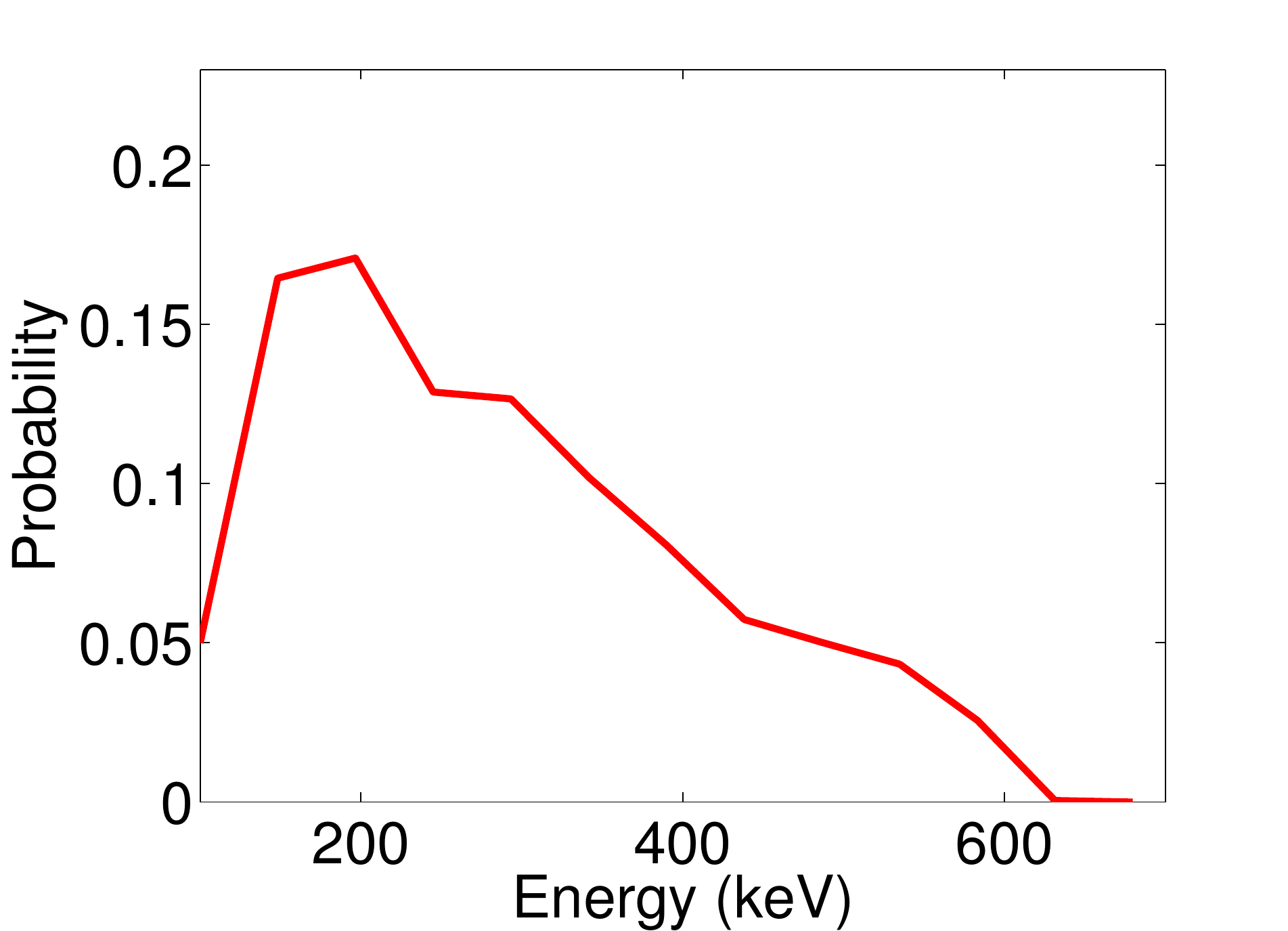}}
	\caption{ Background spectrum created with 2.60\% K40, 3.49 ppm of uranium, and 11.09 ppm of thorium.}
	\label{fig:TucsonBackground}
\end{figure}

The gamma radiation background spectrum was generated using the Gamma Detector and Response Software (GADRAS)\cite{GADRAS}. Because GADRAS does not include liquid scintillators in its list of detector materials, a plastic scintillator was created that best modeled the known physical properties of the liquid scintillator in our detector. Spectral templates for this geometry were created for 1.01\% K40 (from the earth's mantle), 10~ppm of thorium, and 5~ppm of uranium (both from soil).  Using these templates, background spectra can be created for different outdoor locations. An example gamma-ray background is shown in Fig. \ref{fig:TucsonBackground}. This background spectrum was applied to all pixels.

\subsection{Experimental Outline}
For each object, calibration data was simulated for multiple orientations to develop the $H_2$ and $H_1$ probabilities needed to evaluate the observer models. Testing data was independently simulated from another orientation. The observer models acted on this testing data, returning a scalar value.

\subsection{Evaluating Performance}
The metric chosen to evaluate the models was the area under the ROC curve (AUC) \cite{ROC}. We chose this metric because we are not immediately concerned with where to set the test-statistic threshold in the observer studies due to unknown costs associated with incorrect outcomes. Generally, the AUC increases along with the acquisition time. In particular, we will occasionally point out the number of counts corresponding to an AUC of 0.9 to give an easy objective comparison. 

Rather than laboriously generating the ROC curve for each observer, we used the two-alternative forced-choice test (2AFC) ~\cite{2AFC} to calculate the AUC metric. With this method, the observer is presented with a series of pairs of testing datasets. In each pair, one dataset is from a measurement of source 1, and the other is from a measurement of source 2. For each dataset, the observer calculates a test statistic (e.g. $\Lambda$ in \eqref{eqn:IdObNuisanceSKE1}) that is intended to have a higher value for source 2 than for source 1. The AUC is equivalent to the fraction of the time that the source 2 test statistic is greater. In studies that incorporate nuisance parameters, the nuisance parameters are randomly sampled for their respective distributions, data sets are generated from those nuisance parameters, and the 2 AFC test is performed to find the AUC value. For a more detailed discussion of 2AFC studies and a proof that the percentage correct in a 2AFC study is equal to the AUC, we refer the reader to reference \cite{2AFC}.

\section{Experiments and Results}
Performance of the observer models discussed in this section is task-dependent. In this case, distinguishing IO8 and IO9 is a fairly easy task due to differences in their gamma spectra and count rate, as shown in Fig. \ref{fig:IO8v9GammaSpectra}.  It should be noted that the methods used in this paper can be used to study the importance of particular nuisance parameters.  In the following sections, we evaluate the effect that different nuisance parameters have on task performance and how properly accounting for these nuisance parameters can maximize task performance. From this methodology, the host could build a list of nuisance parameters critical to task performance that must be accounted for. Likewise, a performance analysis can demonstrate what nuisance parameters could be ignored.

These simulations were carried out with a few significant assumptions. The background was assumed to be spatially independent on the detector, so each pixel received the same mean count rate and spectra due to the background. Perfect PSD was assumed, meaning there was no misclassification of gammas as neutrons. More complicated detector effects such as pile up or dead time were ignored. Aside from the stated nuisance parameter studied in each model, no further nuisance parameters were assumed. The reason multiple nuisance parameters were not considered in any of the studies is due to computational expediency---the simulation of data sets is a time consuming process, and the number of data sets needed increases exponentially with the number of nuisance parameters. 

The objects were imaged with their vertical axis in construction aligned with the imaging axis. The detector system was simulated with a source to mask distance of \cm{70.5}, mask to detector distance of \cm{60}, mask element size of \cm{1.21} and mask thickness of \cm{6.95}. 

\subsection{SKE Ideal Observer with Varying Background Strength}
This simulation study is analogous to a real world experiment where the monitor is attempting to discriminate IO8 and IO9 with known orientation and location under multiple background strengths. The goal here is to analyze performance of the SKE ideal observer (\eqref{eqn:IdObNuisanceSKE1}) when the testing data is of the same orientation that the observer model was trained on. Calibration data for IO8 and IO9 with Arvo rotation 111 was used to develop the parameters $\overline{N_1}$, $\overline{N_2}$ and the probability densities $pr(A_n|\gamma_1,H_1)$ , $pr(A_n|\gamma_2,H_2)$ in the SKE ideal observer. Testing data for IO8 and IO9 were simulated under the same conditions, and the AUC was found for a given mean number of detected signal counts (proportional to acquisition time). 

\begin{figure}
\centerline{\includegraphics[width=.9\columnwidth]{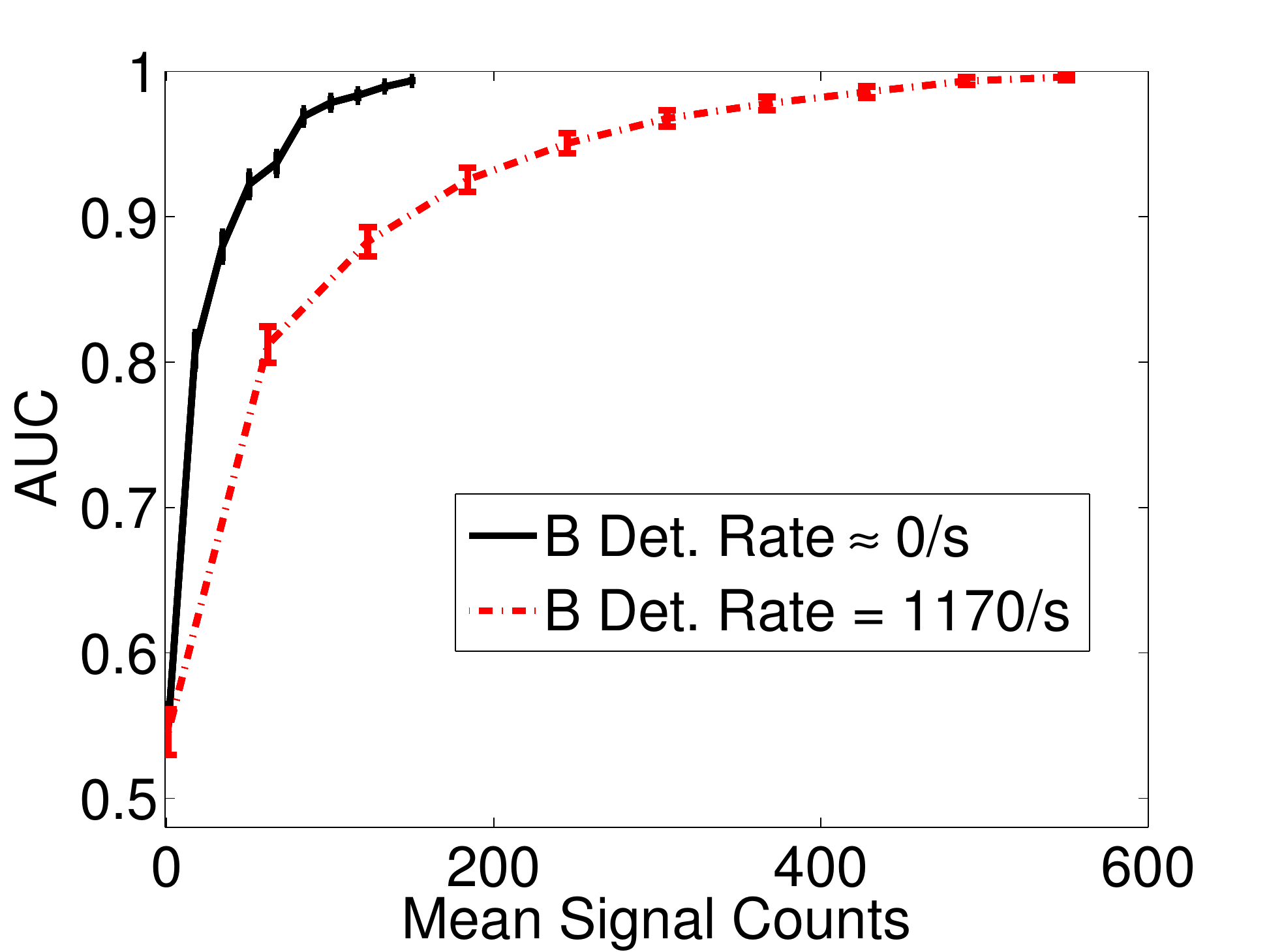}}
\caption{The SKE observer was applied when both the calibration and testing data from sources IO8 and IO9 are from the same orientation. With a weak background, the observer shows extremely good performance. When the background to signal ratio is more realistic, about 4 times as many counts are required for an AUC of 0.9.}
\label{fig:SKEvaryingbackground}
\end{figure}

Performance with a very weak (close to 0 counts per second) and realistic background (1170 counts per second) is shown in Fig. \ref{fig:SKEvaryingbackground}. In the weak background case, only 45 signal counts are required for an AUC of 0.9, while in the realistic background case, 180 signal counts were required. This study shows strong performance of the LM ideal observer under SKE conditions in this task and serves as the baseline for upcoming studies incorporating nuisance parameters.

\subsection{Ideal Observer with Orientation Variability}
Here we assume that objects of an unknown orientation are put inside a container and are imaged by the detector, with every orientation of the source equally likely. The goal of this experiment is to classify sources regardless of their orientation. A total of 60 evenly-spaced orientations of the objects were imaged in accordance with section 3B. The Bayesian prior for the nuisance parameter was built assuming each of these orientations was equally likely. The assumption in this section is that the tested sources have the same pdf on the orientation nuisance parameter as the training sources.  These studies were done with the strong background. We discuss two separate studies.

\begin{figure}
\centerline{\includegraphics[width=.9\columnwidth]{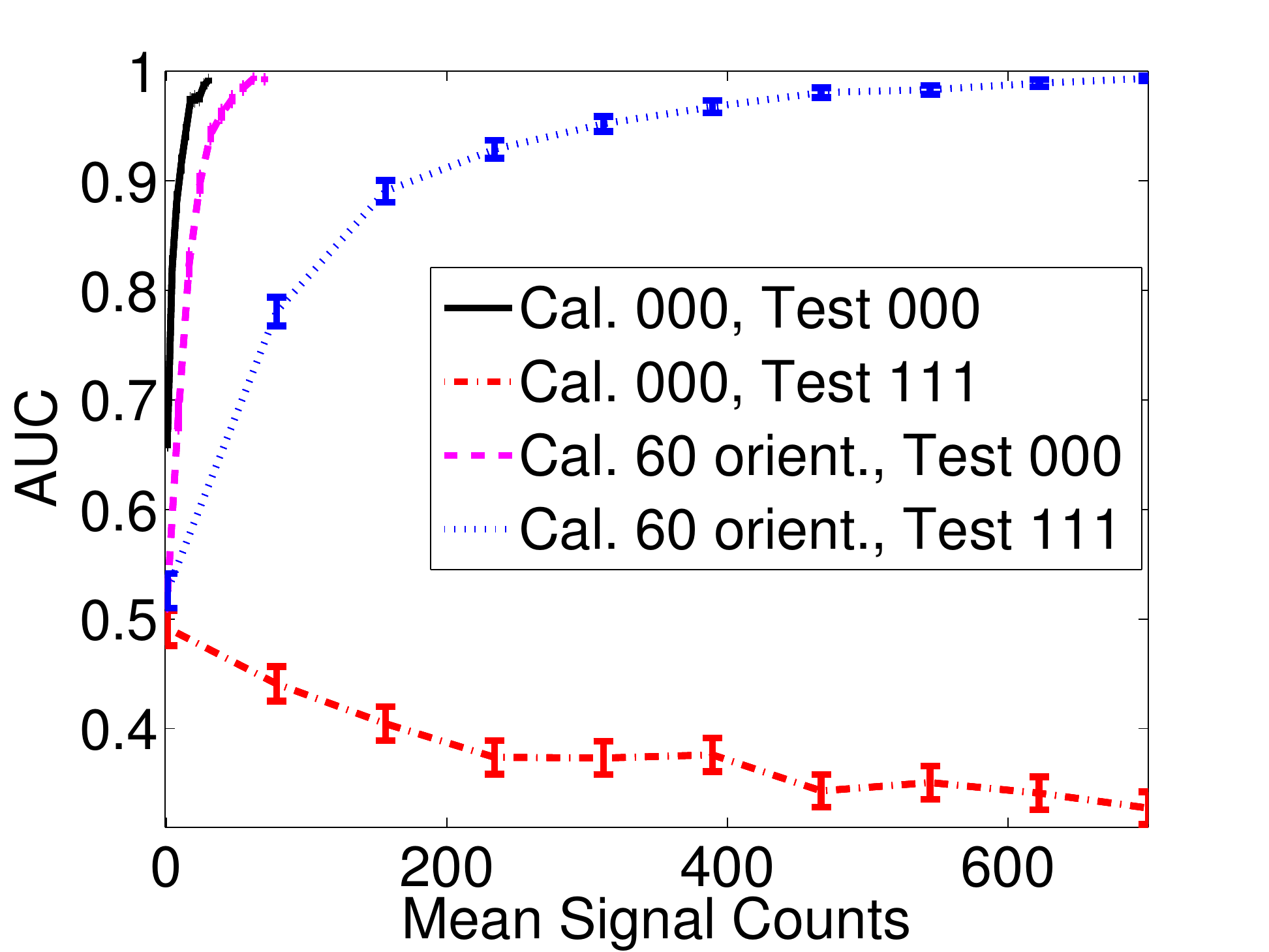}}
\caption{When the testing data is taken from a different orientation than the SKE ideal observer was derived, performance declines. Averaging over the orientation nuisance parameter in the observer model improves performance.
}
\label{fig:Rotation000v111Study}
\end{figure}

The first study (see Fig. \ref{fig:Rotation000v111Study}) highlights the benefits of including the nuisance parameters in the observer model. An SKE ideal observer was found for the simulated sources with Arvo rotation 000. It was used to discriminate IO8 and IO9 testing data using the 2 AFC test for Arvo rotation 000---as opposed to rotation 111 in the prior section---and performance in this task is very strong. This model was then used to classify IO8 and IO9 data under rotation 111, and the observer performs worse than the guessing observer. These two orientations offer the most extreme disparity between the two sources, and predictably performance is quite poor in the case where the observer and testing data are mismatched. The observer model that averaged over all 60 orientations as in \eqref{eqn:IdObNuisance} was then used to discriminate simulated IO8 and IO9 data sets under each orientation. Performance improves upon using the observer that accounts for both orientations when classifying data from the 111 rotation.

\begin{figure}
\centerline{\includegraphics[width=.9\columnwidth]{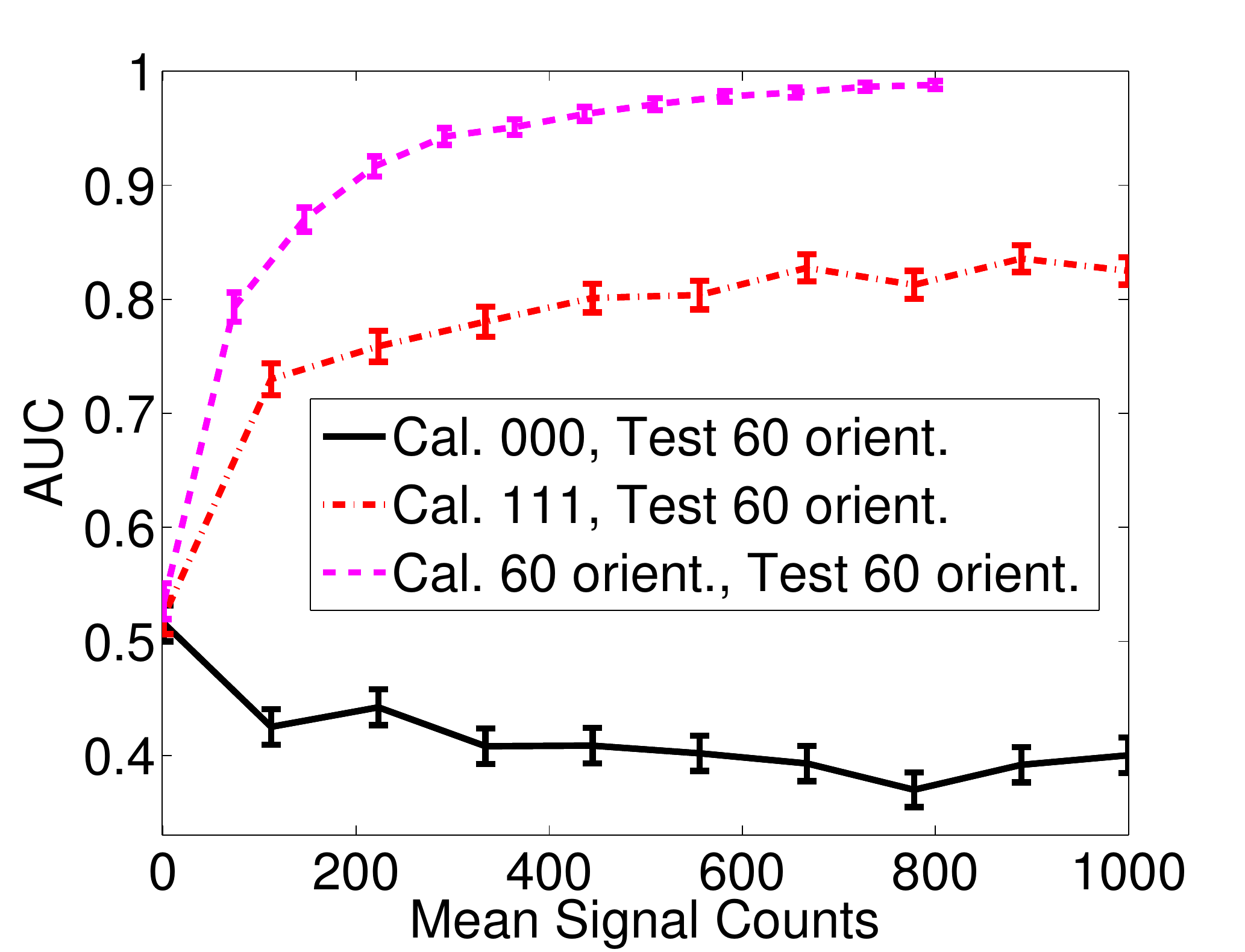}}
\caption{Performance varies significantly when using just a single orientation for IO8 and IO9 calibration data in the observer model to discriminate these sources when testing a random orientation. The observer integrating over all orientations performs the best.
}
\label{fig:RotationStudy}
\end{figure}

In the second study, rather than look at an individual orientation for the testing data, the testing data was randomly sampled from one of the 60 orientations. Here we see that an observer developed from the 000 orientation performs worse than the guessing observer in classifying all of the training data orientations, while the 111 rotation, whose data is more representative of most of the 60 rotations, performs fairly well. As expected, performance is best when the observer model that averages over the orientation nuisance parameter acts on the randomly sampled testing data, as in Fig. \ref{fig:RotationStudy}. This study emphasizes that while strong (but non-optimal) performance can be retained without properly accounting for nuisance parameters, it is subject to the chosen nuisance parameter value or prior density on which the observer is built.

\subsection{Ideal Observer with Count-Rate Variability}
\begin{figure}
\centerline{\includegraphics[width=.9\columnwidth]{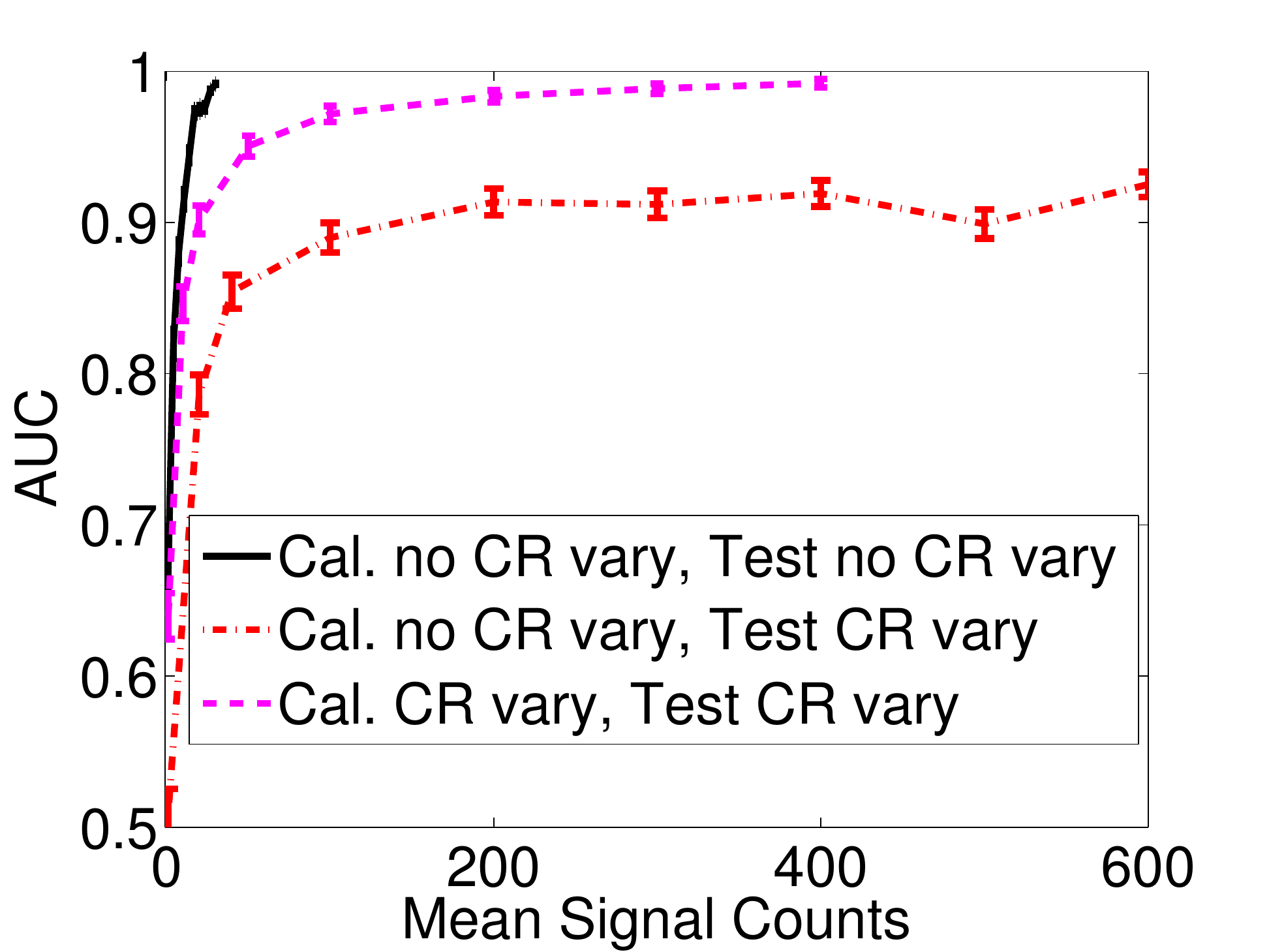}}
\caption{ Performance of the SKE observer (known CR) declines when a range of source activity rates among the testing objects is assumed. Performance improves upon using the ideal observer that averages over the CR nuisance parameter (\eqref{eqn:IdObNuisanceCountRate}) with the correct probability density on the CR. }
\label{fig:VariableCountRateStudy}
\end{figure}

This section presents a practical implementation of the observer model derived in the posterior pdf theory section. This study represents a real-life scenario where a set of sources were created with the same geometry (and thus all should be classified as the same source) but with different emission rates. The observer assumes that there is a spread on activity rates, leading to a probability density over $\overline{N_1}$ and $\overline{N_2}$. A variable background strength is also accounted for in this study, and its corresponding detection rate $\overline{N_b}$ is an example of a shared nuisance parameter $\gamma_0$. The practical implementation of \eqref{eqn:IdObNuisancePost3} in this instance is
\begin{equation}\label{eqn:IdObNuisanceCountRate}
\begin{split}
\Lambda(\{A_n\},N)=&\int\int\int\Lambda_{SKE}(\{A_n\},N|\overline{N_b}, \overline{N_1},\overline{N_2})\\
&pr(\overline{N_b},\overline{N_1}|\{A_n\},N,H_1) pr(\overline{N_2}) \\
&\mathrm{d}\overline{N_b}\mathrm{d}\overline{N_1} \mathrm{d}\overline{N_2}.
\end{split}
\end{equation}
IO8 and IO9 calibration data from Arvo rotation 000 was read in and the observed count rate (CR) was assumed to be the mean of a normal pdf with a standard deviation equal to 40\% of the mean. Sample data was found through randomly sampling the mean number of counts on each pixel according to this posterior density. In Fig. \ref{fig:VariableCountRateStudy}, we see that just using the initial single set of calibration data, the observer model does a poor job classifying IO8 and IO9 objects with varying count rate, with an AUC value that flattens out around 0.9 from 100 to 500 signal counts. Using the observer model that incorporates count-rate variability in the above equation, we are able to achieve better performance.

\section{Conclusion}
This work lays the foundation to identify nuclear sources using LM data in an observer framework. The ideal observer models developed here can be used to incorporate most source-dependent nuisance parameters. Performing binary discrimination under SKE conditions with known values for the nuisance parameters will lead to the best performance, but as object variability is incorporated in testing, performance of the SKE ideal observer declines, sometimes significantly. Including all possible nuisance parameters in these models will lead to the best possible results in performing realistic two-source discrimination problems. However, some nuisance parameters will have a limited or negligible effect on performance in certain tasks, such as the azimuthal angle orientation in a rotationally symmetric object.

While this work provides a foundation for models that use LM data, there is more to study. The ideal observer requires significant storage that necessitates an IB to prevent the monitor from accessing the sensitive calibration data. The monitoring party's inability to access that data would likely make it reluctant to agree to such a model in a treaty-verification setting. We are investigating a linear observer model that penalizes performance based on sensitive information, offering a method to discriminate between objects while not storing information that can be used to reconstruct certain parameters. In addition, binary classification tasks are not ideal for treaty verification. N-source discrimination observers and null hypothesis tests will be studied. 

To accurately predict task performance, the forward model and object scene would also need to be made more physically realistic. The GEANT4 simulations did not include any other objects in the environment such as a floor or walls; hence more scattered particles would be detected in a real-life experiment than these simulations, which would likely make the task more difficult. In addition, imperfect pulse-shape discrimination was ignored here. In practice, photons and neutrons are discriminated based on a measure of their delayed fluorescence \cite{PSD}, but classification is imperfect results in misclassification of neutrons as gamma rays.

\section{Funding}
This work is supported by the Office of Defense Nuclear Nonproliferation Research and Development, Nuclear Weapon and Material Security Team. Sandia National Laboratories is a multi-program laboratory managed and operated by Sandia Corporation, a wholly owned subsidiary of Lockheed Martin Corporation, for the U.S. Department of Energy's National Nuclear Security Administration under contract DE-AC04-94AL85000. (SAND2016-0849J)

Christopher MacGahan was partially funded by the Technology Research Initiative Fund (TRIF) imaging fellowship during the 2013-2014 academic year.

\section{Acknowledgement} 
The authors thank the two anonymous reviewers for their careful reading of this paper and substantive recommendations.

\bibliography{JOSAPaper}

\begin{thebibliography}{27}%
\makeatletter
\providecommand \@ifxundefined [1]{%
 \@ifx{#1\undefined}
}%
\providecommand \@ifnum [1]{%
 \ifnum #1\expandafter \@firstoftwo
 \else \expandafter \@secondoftwo
 \fi
}%
\providecommand \@ifx [1]{%
 \ifx #1\expandafter \@firstoftwo
 \else \expandafter \@secondoftwo
 \fi
}%
\providecommand \natexlab [1]{#1}%
\providecommand \enquote  [1]{``#1''}%
\providecommand \bibnamefont  [1]{#1}%
\providecommand \bibfnamefont [1]{#1}%
\providecommand \citenamefont [1]{#1}%
\providecommand \href@noop [0]{\@secondoftwo}%
\providecommand \href [0]{\begingroup \@sanitize@url \@href}%
\providecommand \@href[1]{\@@startlink{#1}\@@href}%
\providecommand \@@href[1]{\endgroup#1\@@endlink}%
\providecommand \@sanitize@url [0]{\catcode `\\12\catcode `\$12\catcode
  `\&12\catcode `\#12\catcode `\^12\catcode `\_12\catcode `\%12\relax}%
\providecommand \@@startlink[1]{}%
\providecommand \@@endlink[0]{}%
\providecommand \url  [0]{\begingroup\@sanitize@url \@url }%
\providecommand \@url [1]{\endgroup\@href {#1}{\urlprefix }}%
\providecommand \urlprefix  [0]{URL }%
\providecommand \Eprint [0]{\href }%
\providecommand \doibase [0]{http://dx.doi.org/}%
\providecommand \selectlanguage [0]{\@gobble}%
\providecommand \bibinfo  [0]{\@secondoftwo}%
\providecommand \bibfield  [0]{\@secondoftwo}%
\providecommand \translation [1]{[#1]}%
\providecommand \BibitemOpen [0]{}%
\providecommand \bibitemStop [0]{}%
\providecommand \bibitemNoStop [0]{.\EOS\space}%
\providecommand \EOS [0]{\spacefactor3000\relax}%
\providecommand \BibitemShut  [1]{\csname bibitem#1\endcsname}%
\let\auto@bib@innerbib\@empty
\bibitem [{\citenamefont {Zuhoski}\ \emph {et~al.}(1999)\citenamefont
  {Zuhoski}, \citenamefont {Indusi},\ and\ \citenamefont {Vanier}}]{CIVET}%
  \BibitemOpen
  \bibfield  {author} {\bibinfo {author} {\bibfnamefont {P.~B.}\ \bibnamefont
  {Zuhoski}}, \bibinfo {author} {\bibfnamefont {J.~P.}\ \bibnamefont {Indusi}},
  \ and\ \bibinfo {author} {\bibfnamefont {P.~E.}\ \bibnamefont {Vanier}},\
  }\href@noop {} {\emph {\bibinfo {title} {Building a Dedicated Information
  Barrier System for Warhead and Sensitive Item Verification}}},\ \bibinfo
  {type} {Tech. Rep.}\ \bibinfo {number} {BNL-66214}\ (\bibinfo  {institution}
  {Brookhaven National Lab},\ \bibinfo {year} {1999})\BibitemShut {NoStop}%
\bibitem [{\citenamefont {Seager}\ \emph {et~al.}(2001)\citenamefont {Seager},
  \citenamefont {Mitchell}, \citenamefont {Laub}, \citenamefont {Tolk},
  \citenamefont {Lucero},\ and\ \citenamefont {Insch}}]{TRIS}%
  \BibitemOpen
  \bibfield  {author} {\bibinfo {author} {\bibfnamefont {K.}~\bibnamefont
  {Seager}}, \bibinfo {author} {\bibfnamefont {D.}~\bibnamefont {Mitchell}},
  \bibinfo {author} {\bibfnamefont {T.}~\bibnamefont {Laub}}, \bibinfo {author}
  {\bibfnamefont {K.}~\bibnamefont {Tolk}}, \bibinfo {author} {\bibfnamefont
  {R.}~\bibnamefont {Lucero}}, \ and\ \bibinfo {author} {\bibfnamefont
  {K.}~\bibnamefont {Insch}},\ }in\ \href@noop {} {\emph {\bibinfo {booktitle}
  {Proceedings of the 42nd Annual INMM Meeting}}}\ (\bibinfo {address} {Indian
  Wells, CA},\ \bibinfo {year} {2001})\BibitemShut {NoStop}%
\bibitem [{\citenamefont {Mitchell}\ and\ \citenamefont {Tolk}(2000)}]{TRAD}%
  \BibitemOpen
  \bibfield  {author} {\bibinfo {author} {\bibfnamefont {D.}~\bibnamefont
  {Mitchell}}\ and\ \bibinfo {author} {\bibfnamefont {K.}~\bibnamefont
  {Tolk}},\ }in\ \href@noop {} {\emph {\bibinfo {booktitle} {Proceedings of the
  41st Annual INMM Meeting}}}\ (\bibinfo {address} {New Orleans, LA},\ \bibinfo
  {year} {2000})\BibitemShut {NoStop}%
\bibitem [{\citenamefont {Geelhood}\ \emph {et~al.}(2000)\citenamefont
  {Geelhood}, \citenamefont {Bartos}, \citenamefont {Comerford}, \citenamefont
  {Lee}, \citenamefont {Mullens},\ and\ \citenamefont
  {Wolford}}]{TRISTRADREVIEW}%
  \BibitemOpen
  \bibfield  {author} {\bibinfo {author} {\bibfnamefont {B.}~\bibnamefont
  {Geelhood}}, \bibinfo {author} {\bibfnamefont {B.}~\bibnamefont {Bartos}},
  \bibinfo {author} {\bibfnamefont {R.}~\bibnamefont {Comerford}}, \bibinfo
  {author} {\bibfnamefont {D.}~\bibnamefont {Lee}}, \bibinfo {author}
  {\bibfnamefont {J.}~\bibnamefont {Mullens}}, \ and\ \bibinfo {author}
  {\bibfnamefont {J.}~\bibnamefont {Wolford}},\ }\href@noop {} {\emph {\bibinfo
  {title} {Information Barrier Working Group Evaluation of TRADS with
  Particular Attention to Its Authentication Merits}}},\ \bibinfo {type} {Tech.
  Rep.}\ \bibinfo {number} {PNNL-13259}\ (\bibinfo {year} {2000})\BibitemShut
  {NoStop}%
\bibitem [{\citenamefont {Safavian}\ and\ \citenamefont
  {Landgrebe}(1991)}]{TreeSurvey}%
  \BibitemOpen
  \bibfield  {author} {\bibinfo {author} {\bibfnamefont {S.~R.}\ \bibnamefont
  {Safavian}}\ and\ \bibinfo {author} {\bibfnamefont {D.}~\bibnamefont
  {Landgrebe}},\ }\href@noop {} {\bibfield  {journal} {\bibinfo  {journal}
  {IEEE transactions on systems, man, and cybernetics}\ }\textbf {\bibinfo
  {volume} {21}},\ \bibinfo {pages} {660} (\bibinfo {year} {1991})}\BibitemShut
  {NoStop}%
\bibitem [{\citenamefont {Sch{\"o}lkopf}\ and\ \citenamefont
  {Smola}(1998)}]{SVM}%
  \BibitemOpen
  \bibfield  {author} {\bibinfo {author} {\bibfnamefont {B.}~\bibnamefont
  {Sch{\"o}lkopf}}\ and\ \bibinfo {author} {\bibfnamefont {A.}~\bibnamefont
  {Smola}},\ }\href@noop {} {\bibfield  {journal} {\bibinfo  {journal}
  {Encyclopedia of Biostatistics}\ } (\bibinfo {year} {1998})}\BibitemShut
  {NoStop}%
\bibitem [{\citenamefont {Dreiseitl}\ and\ \citenamefont
  {Ohno-Machado}(2002)}]{NNMethodology}%
  \BibitemOpen
  \bibfield  {author} {\bibinfo {author} {\bibfnamefont {S.}~\bibnamefont
  {Dreiseitl}}\ and\ \bibinfo {author} {\bibfnamefont {L.}~\bibnamefont
  {Ohno-Machado}},\ }\href@noop {} {\bibfield  {journal} {\bibinfo  {journal}
  {Journal of biomedical informatics}\ }\textbf {\bibinfo {volume} {35}},\
  \bibinfo {pages} {352} (\bibinfo {year} {2002})}\BibitemShut {NoStop}%
\bibitem [{\citenamefont {Breiman}(2001)}]{RFBreiman}%
  \BibitemOpen
  \bibfield  {author} {\bibinfo {author} {\bibfnamefont {L.}~\bibnamefont
  {Breiman}},\ }\href@noop {} {\bibfield  {journal} {\bibinfo  {journal}
  {Machine learning}\ }\textbf {\bibinfo {volume} {45}},\ \bibinfo {pages} {5}
  (\bibinfo {year} {2001})}\BibitemShut {NoStop}%
\bibitem [{\citenamefont {Liaw}\ and\ \citenamefont {Wiener}(2002)}]{RFLiaw}%
  \BibitemOpen
  \bibfield  {author} {\bibinfo {author} {\bibfnamefont {A.}~\bibnamefont
  {Liaw}}\ and\ \bibinfo {author} {\bibfnamefont {M.}~\bibnamefont {Wiener}},\
  }\href@noop {} {\bibfield  {journal} {\bibinfo  {journal} {R news}\ }\textbf
  {\bibinfo {volume} {2}},\ \bibinfo {pages} {18} (\bibinfo {year}
  {2002})}\BibitemShut {NoStop}%
\bibitem [{\citenamefont {Barrett}\ and\ \citenamefont
  {Myers}(2003)}]{Foundations}%
  \BibitemOpen
  \bibfield  {author} {\bibinfo {author} {\bibfnamefont {H.~H.}\ \bibnamefont
  {Barrett}}\ and\ \bibinfo {author} {\bibfnamefont {K.~J.}\ \bibnamefont
  {Myers}},\ }\enquote {\bibinfo {title} {Foundations of image science},}\ \
  (\bibinfo  {publisher} {Wiley-VCH},\ \bibinfo {year} {2003})\ pp.\ \bibinfo
  {pages} {825--835}\BibitemShut {NoStop}%
\bibitem [{\citenamefont {MacGahan}\ \emph {et~al.}(2014)\citenamefont
  {MacGahan}, \citenamefont {Kupinski}, \citenamefont {Hilton}, \citenamefont
  {Brubaker},\ and\ \citenamefont {Johnson}}]{2014NSSCP}%
  \BibitemOpen
  \bibfield  {author} {\bibinfo {author} {\bibfnamefont {C.~J.}\ \bibnamefont
  {MacGahan}}, \bibinfo {author} {\bibfnamefont {M.~A.}\ \bibnamefont
  {Kupinski}}, \bibinfo {author} {\bibfnamefont {N.~R.}\ \bibnamefont
  {Hilton}}, \bibinfo {author} {\bibfnamefont {E.~M.}\ \bibnamefont
  {Brubaker}}, \ and\ \bibinfo {author} {\bibfnamefont {W.~C.}\ \bibnamefont
  {Johnson}},\ }in\ \href@noop {} {\emph {\bibinfo {booktitle} {Nuclear Science
  Symposium and Medical Imaging Conference (NSS/MIC), 2014 IEEE}}}\ (\bibinfo
  {year} {2014})\ pp.\ \bibinfo {pages} {1--5}\BibitemShut {NoStop}%
\bibitem [{\citenamefont {Barrett}\ \emph {et~al.}(1997)\citenamefont
  {Barrett}, \citenamefont {White},\ and\ \citenamefont {Parra}}]{LMParra}%
  \BibitemOpen
  \bibfield  {author} {\bibinfo {author} {\bibfnamefont {H.~H.}\ \bibnamefont
  {Barrett}}, \bibinfo {author} {\bibfnamefont {T.}~\bibnamefont {White}}, \
  and\ \bibinfo {author} {\bibfnamefont {L.~C.}\ \bibnamefont {Parra}},\
  }\href@noop {} {\bibfield  {journal} {\bibinfo  {journal} {JOSA A}\ }\textbf
  {\bibinfo {volume} {14}},\ \bibinfo {pages} {2914} (\bibinfo {year}
  {1997})}\BibitemShut {NoStop}%
\bibitem [{\citenamefont {Caucci}\ and\ \citenamefont
  {Barrett}(2012)}]{LMCaucciOA}%
  \BibitemOpen
  \bibfield  {author} {\bibinfo {author} {\bibfnamefont {L.}~\bibnamefont
  {Caucci}}\ and\ \bibinfo {author} {\bibfnamefont {H.~H.}\ \bibnamefont
  {Barrett}},\ }\href@noop {} {\bibfield  {journal} {\bibinfo  {journal} {JOSA
  A}\ }\textbf {\bibinfo {volume} {29}},\ \bibinfo {pages} {1003} (\bibinfo
  {year} {2012})}\BibitemShut {NoStop}%
\bibitem [{\citenamefont {Clarkson}(2012)}]{clarkson2012asymptotic}%
  \BibitemOpen
  \bibfield  {author} {\bibinfo {author} {\bibfnamefont {E.}~\bibnamefont
  {Clarkson}},\ }\href@noop {} {\bibfield  {journal} {\bibinfo  {journal} {JOSA
  A}\ }\textbf {\bibinfo {volume} {29}},\ \bibinfo {pages} {2204} (\bibinfo
  {year} {2012})}\BibitemShut {NoStop}%
\bibitem [{\citenamefont {Jha}\ \emph {et~al.}(2013)\citenamefont {Jha},
  \citenamefont {Barrett}, \citenamefont {Clarkson}, \citenamefont {Caucci},\
  and\ \citenamefont {Kupinski}}]{jha2013analytic}%
  \BibitemOpen
  \bibfield  {author} {\bibinfo {author} {\bibfnamefont {A.~K.}\ \bibnamefont
  {Jha}}, \bibinfo {author} {\bibfnamefont {H.~H.}\ \bibnamefont {Barrett}},
  \bibinfo {author} {\bibfnamefont {E.}~\bibnamefont {Clarkson}}, \bibinfo
  {author} {\bibfnamefont {L.}~\bibnamefont {Caucci}}, \ and\ \bibinfo {author}
  {\bibfnamefont {M.~A.}\ \bibnamefont {Kupinski}},\ }\href@noop {} {\bibfield
  {journal} {\bibinfo  {journal} {Intl Meet Fully Three-Dim Image Recon Rad
  Nucl Med, California}\ } (\bibinfo {year} {2013})}\BibitemShut {NoStop}%
\bibitem [{\citenamefont {Asmussen}\ and\ \citenamefont
  {Glynn}(2007)}]{MCSims}%
  \BibitemOpen
  \bibfield  {author} {\bibinfo {author} {\bibfnamefont {S.}~\bibnamefont
  {Asmussen}}\ and\ \bibinfo {author} {\bibfnamefont {P.~W.}\ \bibnamefont
  {Glynn}},\ }\href@noop {} {\emph {\bibinfo {title} {Stochastic simulation:
  Algorithms and analysis}}},\ Vol.~\bibinfo {volume} {57}\ (\bibinfo
  {publisher} {Springer Science \& Business Media},\ \bibinfo {year}
  {2007})\BibitemShut {NoStop}%
\bibitem [{\citenamefont {Gilks}(2005)}]{MCMC}%
  \BibitemOpen
  \bibfield  {author} {\bibinfo {author} {\bibfnamefont {W.~R.}\ \bibnamefont
  {Gilks}},\ }\href@noop {} {\emph {\bibinfo {title} {Markov chain monte
  carlo}}}\ (\bibinfo  {publisher} {Wiley Online Library},\ \bibinfo {year}
  {2005})\BibitemShut {NoStop}%
\bibitem [{\citenamefont {Kupinski}\ \emph {et~al.}(2003)\citenamefont
  {Kupinski}, \citenamefont {Hoppin}, \citenamefont {Clarkson},\ and\
  \citenamefont {Barrett}}]{MCMCKupinski}%
  \BibitemOpen
  \bibfield  {author} {\bibinfo {author} {\bibfnamefont {M.~A.}\ \bibnamefont
  {Kupinski}}, \bibinfo {author} {\bibfnamefont {J.~W.}\ \bibnamefont
  {Hoppin}}, \bibinfo {author} {\bibfnamefont {E.}~\bibnamefont {Clarkson}}, \
  and\ \bibinfo {author} {\bibfnamefont {H.~H.}\ \bibnamefont {Barrett}},\
  }\href@noop {} {\bibfield  {journal} {\bibinfo  {journal} {JOSA A}\ }\textbf
  {\bibinfo {volume} {20}},\ \bibinfo {pages} {430} (\bibinfo {year}
  {2003})}\BibitemShut {NoStop}%
\bibitem [{\citenamefont {Neibert}\ \emph {et~al.}(2010)\citenamefont
  {Neibert}, \citenamefont {Zabriskie}, \citenamefont {Knight},\ and\
  \citenamefont {Jones}}]{INL}%
  \BibitemOpen
  \bibfield  {author} {\bibinfo {author} {\bibfnamefont {R.}~\bibnamefont
  {Neibert}}, \bibinfo {author} {\bibfnamefont {J.}~\bibnamefont {Zabriskie}},
  \bibinfo {author} {\bibfnamefont {C.}~\bibnamefont {Knight}}, \ and\ \bibinfo
  {author} {\bibfnamefont {J.~L.}\ \bibnamefont {Jones}},\ }\href@noop {}
  {\emph {\bibinfo {title} {Passive and Active Radiation Measurements
  Capability at the INL Zero Power Physics Reactor (ZPPR) Facility}}},\
  \bibinfo {type} {Tech. Rep.}\ \bibinfo {number} {INL/EXT-11-20876}\ (\bibinfo
   {institution} {Idaho National Laboratory (INL)},\ \bibinfo {year}
  {2010})\BibitemShut {NoStop}%
\bibitem [{\citenamefont {Hausladen}\ \emph {et~al.}(2012)\citenamefont
  {Hausladen}, \citenamefont {Blackston}, \citenamefont {Brubaker},
  \citenamefont {Chichester}, \citenamefont {Marleau},\ and\ \citenamefont
  {Newby}}]{FNDetector}%
  \BibitemOpen
  \bibfield  {author} {\bibinfo {author} {\bibfnamefont {P.}~\bibnamefont
  {Hausladen}}, \bibinfo {author} {\bibfnamefont {M.~A.}\ \bibnamefont
  {Blackston}}, \bibinfo {author} {\bibfnamefont {E.}~\bibnamefont {Brubaker}},
  \bibinfo {author} {\bibfnamefont {D.}~\bibnamefont {Chichester}}, \bibinfo
  {author} {\bibfnamefont {P.}~\bibnamefont {Marleau}}, \ and\ \bibinfo
  {author} {\bibfnamefont {R.~J.}\ \bibnamefont {Newby}},\ }in\ \href@noop {}
  {\emph {\bibinfo {booktitle} {53rd Annual Meeting of the INMM, Orlando, FL,
  USA}}}\ (\bibinfo {year} {2012})\BibitemShut {NoStop}%
\bibitem [{\citenamefont {Agostinelli}\ \emph {et~al.}(2003)\citenamefont
  {Agostinelli}, \citenamefont {Allison}, \citenamefont {Amako}, \citenamefont
  {Apostolakis}, \citenamefont {Araujo}, \citenamefont {Arce}, \citenamefont
  {Asai}, \citenamefont {Axen}, \citenamefont {Banerjee}, \citenamefont
  {Barrand}, \citenamefont {Behner}, \citenamefont {Bellagamba}, \citenamefont
  {Boundreau}, \citenamefont {Broglia}, \citenamefont {Brunengo}, \citenamefont
  {Burkhardt}, \citenamefont {Chauvie}, \citenamefont {Chuma}, \citenamefont
  {Chytracek}, \citenamefont {Cooperman}, \citenamefont {Cosmo}, \citenamefont
  {Degtyarenko}, \citenamefont {Dell'Acqua}, \citenamefont {Depaolo},
  \citenamefont {Dietrich}, \citenamefont {Enami}, \citenamefont {Feliciello},
  \citenamefont {Ferguson}, \citenamefont {Fesefeldt}, \citenamefont {Folger},
  \citenamefont {Foppiano}, \citenamefont {Forti}, \citenamefont {Garelli},
  \citenamefont {Giani}, \citenamefont {Giannitrapani}, \citenamefont {Gibin},
  \citenamefont {Gomez~Cadenas}, \citenamefont {Gonzalez}, \citenamefont
  {Gracia~Abril}, \citenamefont {Greeniaus}, \citenamefont {Greiner},
  \citenamefont {Grichine}, \citenamefont {Grossheim}, \citenamefont
  {Guatelli}, \citenamefont {Gumpliner}, \citenamefont {Hamatsu}, \citenamefont
  {Hashimoto}, \citenamefont {Hasui}, \citenamefont {Heikkinen}, \citenamefont
  {Howard}, \citenamefont {Ivanchenko}, \citenamefont {Johnson}, \citenamefont
  {Jones}, \citenamefont {Kallenbach}, \citenamefont {Kanaya}, \citenamefont
  {Kawabata}, \citenamefont {Kawabata}, \citenamefont {Kawaguti}, \citenamefont
  {Kelner}, \citenamefont {Kent}, \citenamefont {Kimura}, \citenamefont
  {Kodama}, \citenamefont {Kokoulin}, \citenamefont {Kossov}, \citenamefont
  {Kurashige}, \citenamefont {Lamanna}, \citenamefont {Lampen}, \citenamefont
  {Lara}, \citenamefont {Lefebure}, \citenamefont {Lei}, \citenamefont
  {Liendl}, \citenamefont {Lockman}, \citenamefont {Longo}, \citenamefont
  {Magni}, \citenamefont {Maire}, \citenamefont {Medernach}, \citenamefont
  {Minamimoto}, \citenamefont {More~de Freitas}, \citenamefont {Morita},
  \citenamefont {Murakami}, \citenamefont {Nagamatu}, \citenamefont {Nartallo},
  \citenamefont {Nieminen}, \citenamefont {Nishimura}, \citenamefont {Ohtsubo},
  \citenamefont {Okamura}, \citenamefont {O'Neale}, \citenamefont {Oohata},
  \citenamefont {Paech}, \citenamefont {Perl}, \citenamefont {Pfeiffer},
  \citenamefont {Pia}, \citenamefont {Ranjard}, \citenamefont {Rybin},
  \citenamefont {Sadilov}, \citenamefont {Di~Salvo}, \citenamefont {Santin},
  \citenamefont {Sasaki}, \citenamefont {Savvas}, \citenamefont {Sawada},
  \citenamefont {Scherer}, \citenamefont {Sei}, \citenamefont {Sirotenko},
  \citenamefont {Smith}, \citenamefont {Starkov}, \citenamefont {Stoecker},
  \citenamefont {Sulkimo}, \citenamefont {Takahata}, \citenamefont {Tanaka},
  \citenamefont {Tcherniaev}, \citenamefont {Safai~Tehrani}, \citenamefont
  {Tropeano}, \citenamefont {Truscott}, \citenamefont {Uno}, \citenamefont
  {Urban}, \citenamefont {Urban}, \citenamefont {Verderi}, \citenamefont
  {Walkden}, \citenamefont {Wander}, \citenamefont {Weber}, \citenamefont
  {Wellisch}, \citenamefont {Wenaus}, \citenamefont {Williams}, \citenamefont
  {Wright}, \citenamefont {Yamada}, \citenamefont {Yoshida},\ and\
  \citenamefont {Zschiesche}}]{G4Sim}%
  \BibitemOpen
  \bibfield  {author} {\bibinfo {author} {\bibfnamefont {S.}~\bibnamefont
  {Agostinelli}}, \bibinfo {author} {\bibfnamefont {J.}~\bibnamefont
  {Allison}}, \bibinfo {author} {\bibfnamefont {K.}~\bibnamefont {Amako}},
  \bibinfo {author} {\bibfnamefont {J.}~\bibnamefont {Apostolakis}}, \bibinfo
  {author} {\bibfnamefont {H.}~\bibnamefont {Araujo}}, \bibinfo {author}
  {\bibfnamefont {P.}~\bibnamefont {Arce}}, \bibinfo {author} {\bibfnamefont
  {M.}~\bibnamefont {Asai}}, \bibinfo {author} {\bibfnamefont {D.}~\bibnamefont
  {Axen}}, \bibinfo {author} {\bibfnamefont {S.}~\bibnamefont {Banerjee}},
  \bibinfo {author} {\bibfnamefont {G.}~\bibnamefont {Barrand}}, \bibinfo
  {author} {\bibfnamefont {F.}~\bibnamefont {Behner}}, \bibinfo {author}
  {\bibfnamefont {L.}~\bibnamefont {Bellagamba}}, \bibinfo {author}
  {\bibfnamefont {J.}~\bibnamefont {Boundreau}}, \bibinfo {author}
  {\bibfnamefont {L.}~\bibnamefont {Broglia}}, \bibinfo {author} {\bibfnamefont
  {A.}~\bibnamefont {Brunengo}}, \bibinfo {author} {\bibfnamefont
  {H.}~\bibnamefont {Burkhardt}}, \bibinfo {author} {\bibfnamefont
  {S.}~\bibnamefont {Chauvie}}, \bibinfo {author} {\bibfnamefont
  {J.}~\bibnamefont {Chuma}}, \bibinfo {author} {\bibfnamefont
  {R.}~\bibnamefont {Chytracek}}, \bibinfo {author} {\bibfnamefont
  {G.}~\bibnamefont {Cooperman}}, \bibinfo {author} {\bibfnamefont
  {G.}~\bibnamefont {Cosmo}}, \bibinfo {author} {\bibfnamefont
  {P.}~\bibnamefont {Degtyarenko}}, \bibinfo {author} {\bibfnamefont
  {A.}~\bibnamefont {Dell'Acqua}}, \bibinfo {author} {\bibfnamefont
  {G.}~\bibnamefont {Depaolo}}, \bibinfo {author} {\bibfnamefont
  {D.}~\bibnamefont {Dietrich}}, \bibinfo {author} {\bibfnamefont
  {R.}~\bibnamefont {Enami}}, \bibinfo {author} {\bibfnamefont
  {A.}~\bibnamefont {Feliciello}}, \bibinfo {author} {\bibfnamefont
  {C.}~\bibnamefont {Ferguson}}, \bibinfo {author} {\bibfnamefont
  {H.}~\bibnamefont {Fesefeldt}}, \bibinfo {author} {\bibfnamefont
  {G.}~\bibnamefont {Folger}}, \bibinfo {author} {\bibfnamefont
  {F.}~\bibnamefont {Foppiano}}, \bibinfo {author} {\bibfnamefont
  {A.}~\bibnamefont {Forti}}, \bibinfo {author} {\bibfnamefont
  {S.}~\bibnamefont {Garelli}}, \bibinfo {author} {\bibfnamefont
  {S.}~\bibnamefont {Giani}}, \bibinfo {author} {\bibfnamefont
  {R.}~\bibnamefont {Giannitrapani}}, \bibinfo {author} {\bibfnamefont
  {D.}~\bibnamefont {Gibin}}, \bibinfo {author} {\bibfnamefont
  {J.}~\bibnamefont {Gomez~Cadenas}}, \bibinfo {author} {\bibfnamefont
  {I.}~\bibnamefont {Gonzalez}}, \bibinfo {author} {\bibfnamefont
  {G.}~\bibnamefont {Gracia~Abril}}, \bibinfo {author} {\bibfnamefont
  {G.}~\bibnamefont {Greeniaus}}, \bibinfo {author} {\bibfnamefont
  {W.}~\bibnamefont {Greiner}}, \bibinfo {author} {\bibfnamefont
  {V.}~\bibnamefont {Grichine}}, \bibinfo {author} {\bibfnamefont
  {A.}~\bibnamefont {Grossheim}}, \bibinfo {author} {\bibfnamefont
  {S.}~\bibnamefont {Guatelli}}, \bibinfo {author} {\bibfnamefont
  {P.}~\bibnamefont {Gumpliner}}, \bibinfo {author} {\bibfnamefont
  {R.}~\bibnamefont {Hamatsu}}, \bibinfo {author} {\bibfnamefont
  {K.}~\bibnamefont {Hashimoto}}, \bibinfo {author} {\bibfnamefont
  {H.}~\bibnamefont {Hasui}}, \bibinfo {author} {\bibfnamefont
  {A.}~\bibnamefont {Heikkinen}}, \bibinfo {author} {\bibfnamefont
  {A.}~\bibnamefont {Howard}}, \bibinfo {author} {\bibfnamefont
  {V.}~\bibnamefont {Ivanchenko}}, \bibinfo {author} {\bibfnamefont
  {A.}~\bibnamefont {Johnson}}, \bibinfo {author} {\bibfnamefont
  {F.}~\bibnamefont {Jones}}, \bibinfo {author} {\bibfnamefont
  {J.}~\bibnamefont {Kallenbach}}, \bibinfo {author} {\bibfnamefont
  {N.}~\bibnamefont {Kanaya}}, \bibinfo {author} {\bibfnamefont
  {M.}~\bibnamefont {Kawabata}}, \bibinfo {author} {\bibfnamefont
  {Y.}~\bibnamefont {Kawabata}}, \bibinfo {author} {\bibfnamefont
  {M.}~\bibnamefont {Kawaguti}}, \bibinfo {author} {\bibfnamefont
  {S.}~\bibnamefont {Kelner}}, \bibinfo {author} {\bibfnamefont
  {P.}~\bibnamefont {Kent}}, \bibinfo {author} {\bibfnamefont {A.}~\bibnamefont
  {Kimura}}, \bibinfo {author} {\bibfnamefont {T.}~\bibnamefont {Kodama}},
  \bibinfo {author} {\bibfnamefont {R.}~\bibnamefont {Kokoulin}}, \bibinfo
  {author} {\bibfnamefont {M.}~\bibnamefont {Kossov}}, \bibinfo {author}
  {\bibfnamefont {H.}~\bibnamefont {Kurashige}}, \bibinfo {author}
  {\bibfnamefont {E.}~\bibnamefont {Lamanna}}, \bibinfo {author} {\bibfnamefont
  {T.}~\bibnamefont {Lampen}}, \bibinfo {author} {\bibfnamefont
  {V.}~\bibnamefont {Lara}}, \bibinfo {author} {\bibfnamefont {V.}~\bibnamefont
  {Lefebure}}, \bibinfo {author} {\bibfnamefont {F.}~\bibnamefont {Lei}},
  \bibinfo {author} {\bibfnamefont {M.}~\bibnamefont {Liendl}}, \bibinfo
  {author} {\bibfnamefont {W.}~\bibnamefont {Lockman}}, \bibinfo {author}
  {\bibfnamefont {F.}~\bibnamefont {Longo}}, \bibinfo {author} {\bibfnamefont
  {S.}~\bibnamefont {Magni}}, \bibinfo {author} {\bibfnamefont
  {M.}~\bibnamefont {Maire}}, \bibinfo {author} {\bibfnamefont
  {E.}~\bibnamefont {Medernach}}, \bibinfo {author} {\bibfnamefont
  {K.}~\bibnamefont {Minamimoto}}, \bibinfo {author} {\bibfnamefont
  {P.}~\bibnamefont {More~de Freitas}}, \bibinfo {author} {\bibfnamefont
  {Y.}~\bibnamefont {Morita}}, \bibinfo {author} {\bibfnamefont
  {K.}~\bibnamefont {Murakami}}, \bibinfo {author} {\bibfnamefont
  {M.}~\bibnamefont {Nagamatu}}, \bibinfo {author} {\bibfnamefont
  {R.}~\bibnamefont {Nartallo}}, \bibinfo {author} {\bibfnamefont
  {P.}~\bibnamefont {Nieminen}}, \bibinfo {author} {\bibfnamefont
  {T.}~\bibnamefont {Nishimura}}, \bibinfo {author} {\bibfnamefont
  {K.}~\bibnamefont {Ohtsubo}}, \bibinfo {author} {\bibfnamefont
  {M.}~\bibnamefont {Okamura}}, \bibinfo {author} {\bibfnamefont
  {S.}~\bibnamefont {O'Neale}}, \bibinfo {author} {\bibfnamefont
  {Y.}~\bibnamefont {Oohata}}, \bibinfo {author} {\bibfnamefont
  {K.}~\bibnamefont {Paech}}, \bibinfo {author} {\bibfnamefont
  {J.}~\bibnamefont {Perl}}, \bibinfo {author} {\bibfnamefont {A.}~\bibnamefont
  {Pfeiffer}}, \bibinfo {author} {\bibfnamefont {M.}~\bibnamefont {Pia}},
  \bibinfo {author} {\bibfnamefont {F.}~\bibnamefont {Ranjard}}, \bibinfo
  {author} {\bibfnamefont {A.}~\bibnamefont {Rybin}}, \bibinfo {author}
  {\bibfnamefont {S.}~\bibnamefont {Sadilov}}, \bibinfo {author} {\bibfnamefont
  {E.}~\bibnamefont {Di~Salvo}}, \bibinfo {author} {\bibfnamefont
  {G.}~\bibnamefont {Santin}}, \bibinfo {author} {\bibfnamefont
  {T.}~\bibnamefont {Sasaki}}, \bibinfo {author} {\bibfnamefont
  {N.}~\bibnamefont {Savvas}}, \bibinfo {author} {\bibfnamefont
  {Y.}~\bibnamefont {Sawada}}, \bibinfo {author} {\bibfnamefont
  {S.}~\bibnamefont {Scherer}}, \bibinfo {author} {\bibfnamefont
  {S.}~\bibnamefont {Sei}}, \bibinfo {author} {\bibfnamefont {V.}~\bibnamefont
  {Sirotenko}}, \bibinfo {author} {\bibfnamefont {D.}~\bibnamefont {Smith}},
  \bibinfo {author} {\bibfnamefont {N.}~\bibnamefont {Starkov}}, \bibinfo
  {author} {\bibfnamefont {H.}~\bibnamefont {Stoecker}}, \bibinfo {author}
  {\bibfnamefont {J.}~\bibnamefont {Sulkimo}}, \bibinfo {author} {\bibfnamefont
  {M.}~\bibnamefont {Takahata}}, \bibinfo {author} {\bibfnamefont
  {S.}~\bibnamefont {Tanaka}}, \bibinfo {author} {\bibfnamefont
  {E.}~\bibnamefont {Tcherniaev}}, \bibinfo {author} {\bibfnamefont
  {E.}~\bibnamefont {Safai~Tehrani}}, \bibinfo {author} {\bibfnamefont
  {M.}~\bibnamefont {Tropeano}}, \bibinfo {author} {\bibfnamefont
  {P.}~\bibnamefont {Truscott}}, \bibinfo {author} {\bibfnamefont
  {H.}~\bibnamefont {Uno}}, \bibinfo {author} {\bibfnamefont {L.}~\bibnamefont
  {Urban}}, \bibinfo {author} {\bibfnamefont {P.}~\bibnamefont {Urban}},
  \bibinfo {author} {\bibfnamefont {M.}~\bibnamefont {Verderi}}, \bibinfo
  {author} {\bibfnamefont {A.}~\bibnamefont {Walkden}}, \bibinfo {author}
  {\bibfnamefont {W.}~\bibnamefont {Wander}}, \bibinfo {author} {\bibfnamefont
  {H.}~\bibnamefont {Weber}}, \bibinfo {author} {\bibfnamefont
  {J.}~\bibnamefont {Wellisch}}, \bibinfo {author} {\bibfnamefont
  {T.}~\bibnamefont {Wenaus}}, \bibinfo {author} {\bibfnamefont
  {D.}~\bibnamefont {Williams}}, \bibinfo {author} {\bibfnamefont
  {D.}~\bibnamefont {Wright}}, \bibinfo {author} {\bibfnamefont
  {T.}~\bibnamefont {Yamada}}, \bibinfo {author} {\bibfnamefont
  {H.}~\bibnamefont {Yoshida}}, \ and\ \bibinfo {author} {\bibfnamefont
  {D.}~\bibnamefont {Zschiesche}},\ }\href@noop {} {\bibfield  {journal}
  {\bibinfo  {journal} {Nuclear instruments and methods in physics research
  section A: Accelerators, Spectrometers, Detectors and Associated Equipment}\
  }\textbf {\bibinfo {volume} {506}},\ \bibinfo {pages} {250} (\bibinfo {year}
  {2003})}\BibitemShut {NoStop}%
\bibitem [{\citenamefont {Allison}\ \emph {et~al.}(2006)\citenamefont
  {Allison}, \citenamefont {Amako}, \citenamefont {Apostolakis}, \citenamefont
  {Araujo}, \citenamefont {Arce~Dubois}, \citenamefont {Asai}, \citenamefont
  {Barrand}, \citenamefont {Capra}, \citenamefont {Chauvie}, \citenamefont
  {Chytracek}, \citenamefont {Cirrone}, \citenamefont {Cooperman},
  \citenamefont {Cosmo}, \citenamefont {Cuttone}, \citenamefont {Daquino},
  \citenamefont {Donszelmann}, \citenamefont {Dressel}, \citenamefont {Folger},
  \citenamefont {Foppiano}, \citenamefont {Generowicz}, \citenamefont
  {Grichine}, \citenamefont {Guatelli}, \citenamefont {P}, \citenamefont
  {Heikkinen}, \citenamefont {Hrivnacova}, \citenamefont {Howard},
  \citenamefont {Incerti}, \citenamefont {Ivanchenko}, \citenamefont {Johnson},
  \citenamefont {Jones}, \citenamefont {Koi}, \citenamefont {Kokoulin},
  \citenamefont {Kossov}, \citenamefont {Kurashige}, \citenamefont {Lara},
  \citenamefont {Larsson}, \citenamefont {Lei}, \citenamefont {Link},
  \citenamefont {Longo}, \citenamefont {Maire}, \citenamefont {Mantero},
  \citenamefont {Mascialino}, \citenamefont {McLaren}, \citenamefont
  {Mendez~Lorenzo}, \citenamefont {Minamimoto}, \citenamefont {Murakami},
  \citenamefont {Nieminen}, \citenamefont {Pandola}, \citenamefont {Parlati},
  \citenamefont {Peralta}, \citenamefont {Perl}, \citenamefont {Pfeiffer},
  \citenamefont {Pia}, \citenamefont {Ribon}, \citenamefont {Rodrigues},
  \citenamefont {Russo}, \citenamefont {Sadilov}, \citenamefont {Santin},
  \citenamefont {Sasaki}, \citenamefont {Smith}, \citenamefont {Starkov},
  \citenamefont {Tanaka}, \citenamefont {Tcherniaev}, \citenamefont {Tome},
  \citenamefont {Trindade}, \citenamefont {Truscott}, \citenamefont {Urban},
  \citenamefont {Verderi}, \citenamefont {Walkden}, \citenamefont {Wellisch},
  \citenamefont {Williams}, \citenamefont {Wright},\ and\ \citenamefont
  {Yoshida}}]{G4Dev}%
  \BibitemOpen
  \bibfield  {author} {\bibinfo {author} {\bibfnamefont {J.}~\bibnamefont
  {Allison}}, \bibinfo {author} {\bibfnamefont {K.}~\bibnamefont {Amako}},
  \bibinfo {author} {\bibfnamefont {J.}~\bibnamefont {Apostolakis}}, \bibinfo
  {author} {\bibfnamefont {H.}~\bibnamefont {Araujo}}, \bibinfo {author}
  {\bibfnamefont {P.}~\bibnamefont {Arce~Dubois}}, \bibinfo {author}
  {\bibfnamefont {M.}~\bibnamefont {Asai}}, \bibinfo {author} {\bibfnamefont
  {G.}~\bibnamefont {Barrand}}, \bibinfo {author} {\bibfnamefont
  {R.}~\bibnamefont {Capra}}, \bibinfo {author} {\bibfnamefont
  {S.}~\bibnamefont {Chauvie}}, \bibinfo {author} {\bibfnamefont
  {R.}~\bibnamefont {Chytracek}}, \bibinfo {author} {\bibfnamefont
  {G.}~\bibnamefont {Cirrone}}, \bibinfo {author} {\bibfnamefont
  {G.}~\bibnamefont {Cooperman}}, \bibinfo {author} {\bibfnamefont
  {G.}~\bibnamefont {Cosmo}}, \bibinfo {author} {\bibfnamefont
  {G.}~\bibnamefont {Cuttone}}, \bibinfo {author} {\bibfnamefont
  {G.}~\bibnamefont {Daquino}}, \bibinfo {author} {\bibfnamefont
  {M.}~\bibnamefont {Donszelmann}}, \bibinfo {author} {\bibfnamefont
  {M.}~\bibnamefont {Dressel}}, \bibinfo {author} {\bibfnamefont
  {G.}~\bibnamefont {Folger}}, \bibinfo {author} {\bibfnamefont
  {F.}~\bibnamefont {Foppiano}}, \bibinfo {author} {\bibfnamefont
  {J.}~\bibnamefont {Generowicz}}, \bibinfo {author} {\bibfnamefont
  {V.}~\bibnamefont {Grichine}}, \bibinfo {author} {\bibfnamefont
  {S.}~\bibnamefont {Guatelli}}, \bibinfo {author} {\bibfnamefont
  {G.}~\bibnamefont {P}}, \bibinfo {author} {\bibfnamefont {A.}~\bibnamefont
  {Heikkinen}}, \bibinfo {author} {\bibfnamefont {I.}~\bibnamefont
  {Hrivnacova}}, \bibinfo {author} {\bibfnamefont {A.}~\bibnamefont {Howard}},
  \bibinfo {author} {\bibfnamefont {S.}~\bibnamefont {Incerti}}, \bibinfo
  {author} {\bibfnamefont {V.}~\bibnamefont {Ivanchenko}}, \bibinfo {author}
  {\bibfnamefont {T.}~\bibnamefont {Johnson}}, \bibinfo {author} {\bibfnamefont
  {F.}~\bibnamefont {Jones}}, \bibinfo {author} {\bibfnamefont
  {T.}~\bibnamefont {Koi}}, \bibinfo {author} {\bibfnamefont {R.}~\bibnamefont
  {Kokoulin}}, \bibinfo {author} {\bibfnamefont {M.}~\bibnamefont {Kossov}},
  \bibinfo {author} {\bibfnamefont {H.}~\bibnamefont {Kurashige}}, \bibinfo
  {author} {\bibfnamefont {V.}~\bibnamefont {Lara}}, \bibinfo {author}
  {\bibfnamefont {S.}~\bibnamefont {Larsson}}, \bibinfo {author} {\bibfnamefont
  {F.}~\bibnamefont {Lei}}, \bibinfo {author} {\bibfnamefont {O.}~\bibnamefont
  {Link}}, \bibinfo {author} {\bibfnamefont {F.}~\bibnamefont {Longo}},
  \bibinfo {author} {\bibfnamefont {M.}~\bibnamefont {Maire}}, \bibinfo
  {author} {\bibfnamefont {A.}~\bibnamefont {Mantero}}, \bibinfo {author}
  {\bibfnamefont {B.}~\bibnamefont {Mascialino}}, \bibinfo {author}
  {\bibfnamefont {I.}~\bibnamefont {McLaren}}, \bibinfo {author} {\bibfnamefont
  {P.}~\bibnamefont {Mendez~Lorenzo}}, \bibinfo {author} {\bibfnamefont
  {K.}~\bibnamefont {Minamimoto}}, \bibinfo {author} {\bibfnamefont
  {K.}~\bibnamefont {Murakami}}, \bibinfo {author} {\bibfnamefont
  {P.}~\bibnamefont {Nieminen}}, \bibinfo {author} {\bibfnamefont
  {L.}~\bibnamefont {Pandola}}, \bibinfo {author} {\bibfnamefont
  {S.}~\bibnamefont {Parlati}}, \bibinfo {author} {\bibfnamefont
  {L.}~\bibnamefont {Peralta}}, \bibinfo {author} {\bibfnamefont
  {J.}~\bibnamefont {Perl}}, \bibinfo {author} {\bibfnamefont {A.}~\bibnamefont
  {Pfeiffer}}, \bibinfo {author} {\bibfnamefont {M.}~\bibnamefont {Pia}},
  \bibinfo {author} {\bibfnamefont {A.}~\bibnamefont {Ribon}}, \bibinfo
  {author} {\bibfnamefont {P.}~\bibnamefont {Rodrigues}}, \bibinfo {author}
  {\bibfnamefont {G.}~\bibnamefont {Russo}}, \bibinfo {author} {\bibfnamefont
  {S.}~\bibnamefont {Sadilov}}, \bibinfo {author} {\bibfnamefont
  {G.}~\bibnamefont {Santin}}, \bibinfo {author} {\bibfnamefont
  {T.}~\bibnamefont {Sasaki}}, \bibinfo {author} {\bibfnamefont
  {D.}~\bibnamefont {Smith}}, \bibinfo {author} {\bibfnamefont
  {N.}~\bibnamefont {Starkov}}, \bibinfo {author} {\bibfnamefont
  {S.}~\bibnamefont {Tanaka}}, \bibinfo {author} {\bibfnamefont
  {E.}~\bibnamefont {Tcherniaev}}, \bibinfo {author} {\bibfnamefont
  {B.}~\bibnamefont {Tome}}, \bibinfo {author} {\bibfnamefont {A.}~\bibnamefont
  {Trindade}}, \bibinfo {author} {\bibfnamefont {P.}~\bibnamefont {Truscott}},
  \bibinfo {author} {\bibfnamefont {L.}~\bibnamefont {Urban}}, \bibinfo
  {author} {\bibfnamefont {M.}~\bibnamefont {Verderi}}, \bibinfo {author}
  {\bibfnamefont {A.}~\bibnamefont {Walkden}}, \bibinfo {author} {\bibfnamefont
  {J.}~\bibnamefont {Wellisch}}, \bibinfo {author} {\bibfnamefont
  {D.}~\bibnamefont {Williams}}, \bibinfo {author} {\bibfnamefont
  {D.}~\bibnamefont {Wright}}, \ and\ \bibinfo {author} {\bibfnamefont
  {H.}~\bibnamefont {Yoshida}},\ }\href@noop {} {\bibfield  {journal} {\bibinfo
   {journal} {Nuclear Science, IEEE Transactions on}\ }\textbf {\bibinfo
  {volume} {53}},\ \bibinfo {pages} {270} (\bibinfo {year} {2006})}\BibitemShut
  {NoStop}%
\bibitem [{\citenamefont {Arvo}(1992)}]{Arvorotation}%
  \BibitemOpen
  \bibfield  {author} {\bibinfo {author} {\bibfnamefont {J.}~\bibnamefont
  {Arvo}},\ }in\ \href@noop {} {\emph {\bibinfo {booktitle} {Graphics Gems
  III}}}\ (\bibinfo  {publisher} {Academic Press},\ \bibinfo {year} {1992})\
  pp.\ \bibinfo {pages} {117--120}\BibitemShut {NoStop}%
\bibitem [{\citenamefont {Mitchell}(1988)}]{GADRAS}%
  \BibitemOpen
  \bibfield  {author} {\bibinfo {author} {\bibfnamefont {D.~J.}\ \bibnamefont
  {Mitchell}},\ }\href@noop {} {\emph {\bibinfo {title} {Gamma Detector
  Response and Analysis Software (GADRAS)}}},\ \bibinfo {type} {Tech. Rep.}\
  \bibinfo {number} {SAND88-2519}\ (\bibinfo  {institution} {Sandia National
  Laboratories},\ \bibinfo {year} {1988})\BibitemShut {NoStop}%
\bibitem [{\citenamefont {Hanley}\ and\ \citenamefont {McNeil}(1982)}]{ROC}%
  \BibitemOpen
  \bibfield  {author} {\bibinfo {author} {\bibfnamefont {J.~A.}\ \bibnamefont
  {Hanley}}\ and\ \bibinfo {author} {\bibfnamefont {B.~J.}\ \bibnamefont
  {McNeil}},\ }\href@noop {} {\bibfield  {journal} {\bibinfo  {journal}
  {Radiology}\ }\textbf {\bibinfo {volume} {143}},\ \bibinfo {pages} {29}
  (\bibinfo {year} {1982})}\BibitemShut {NoStop}%
\bibitem [{\citenamefont {Fechner}\ \emph {et~al.}(1966)\citenamefont
  {Fechner}, \citenamefont {Boring}, \citenamefont {Howes},\ and\ \citenamefont
  {Adler}}]{2AFC}%
  \BibitemOpen
  \bibfield  {author} {\bibinfo {author} {\bibfnamefont {G.~T.}\ \bibnamefont
  {Fechner}}, \bibinfo {author} {\bibfnamefont {E.~G.}\ \bibnamefont {Boring}},
  \bibinfo {author} {\bibfnamefont {D.~H.}\ \bibnamefont {Howes}}, \ and\
  \bibinfo {author} {\bibfnamefont {H.~E.}\ \bibnamefont {Adler}},\ }\href@noop
  {} {\emph {\bibinfo {title} {Elements of Psychophysics. Translated by Helmut
  E. Adler. Edited by Davis H. Howes And Edwin G. Boring, With an Introd. by
  Edwin G. Boring}}}\ (\bibinfo  {publisher} {Holt, Rinehart and Winston},\
  \bibinfo {year} {1966})\BibitemShut {NoStop}%
\bibitem [{\citenamefont {Adams}\ and\ \citenamefont {White}(1978)}]{PSD}%
  \BibitemOpen
  \bibfield  {author} {\bibinfo {author} {\bibfnamefont {J.}~\bibnamefont
  {Adams}}\ and\ \bibinfo {author} {\bibfnamefont {G.}~\bibnamefont {White}},\
  }\href@noop {} {\bibfield  {journal} {\bibinfo  {journal} {Nuclear
  Instruments and Methods}\ }\textbf {\bibinfo {volume} {156}},\ \bibinfo
  {pages} {459} (\bibinfo {year} {1978})}\BibitemShut {NoStop}%
\end{thebibliography}%

\end{document}